\documentclass{JFM-FLM_Au}

\usepackage{siunitx}
\usepackage{longtable}
\usepackage{float}
\usepackage{subcaption}
\usepackage{multirow}
\usepackage{bm}
\usepackage{cite}

\DeclareMathSymbol{\varOmega}{\mathalpha}{operators}{"0A}

\lefttitle{Christoph Wilms and Holger Grosshans}
\righttitle{Journal of Fluid Mechanics}

\title{Effect of Reynolds number on triboelectric particle charging in turbulent channel flow}

\author{Christoph Wilms\aff{1,2} \and Holger Grosshans\aff{1,2}}

\affiliation{\aff{1}Physikalisch-Technische Bundesanstalt, Bundesallee 100, Braunschweig, Germany
\aff{2}Otto von Guericke University Magdeburg, Universitätsplatz 2, Magdeburg, Germany}

\corresau{Christoph Wilms, \email{Christoph.Wilms@ptb.de}}

\begin{document}
\maketitle

\begin{abstract}
  Triboelectric charging in particle-laden flows is a complex interplay of fluid and particle dynamics, collision mechanics, and electrostatics. In this study, we introduce triboFoam, an open-source solver built on the OpenFOAM framework, designed to simulate triboelectric charging in particle-laden turbulent flows. We validate triboFoam using Direct Numerical Simulations (DNS) of a fully developed turbulent channel flow at a friction Reynolds number of \(\Rey_\tau = 180\). The results demonstrate good agreement with DNS data for particle concentration profiles and charge distributions. Then, we investigate the influence of Reynolds number on particle distribution and charging behaviour using Large-Eddy Simulations (LES) at varying friction Reynolds numbers up to \(\Rey_\tau = 550\). Our findings reveal that higher Reynolds numbers lead to increased near-wall particle concentrations and enhanced charging rates, attributed to intensified turbulent fluctuations and elevated impact velocities. Finally, an empirical correlation is proposed to predict the average particle charging rate as a function of Reynolds number and particle diameter. With this work, we provide a tool for simulating triboelectric charging in complex geometries and turbulent flows, advancing the understanding of electrostatic phenomena in particle-laden systems. The empirical correlation offers practical insights for predicting charging behaviour in industrial applications and thus can contribute to improved safety and efficiency in processes involving particulate matter.
\end{abstract}



\section{Introduction}\label{introduction}

During the handling of polymers, pharmaceuticals, food, and other powdered materials, particles frequently become electrified via triboelectric (frictional) charging.
Electrostatic charge can result in many undesirable effects, such as particle segregation, adhesion, fouling, blockages of pipes, reduced system efficiency, and even dust explosions~\citep{Giffin2013,Osh11}.
Many industrial dust explosions have been attributed to the discharge of static electricity, for example, when filling silos with fine powders and high particle throughput~\citep{Glor2001, Nifuku2003,ZHANG26}.
Among all powder handling operations, including sieving, pouring, and grinding, the highest charge appears in pneumatic conveying, where the particles are transported by a highly turbulent flow~\citep{Glor1985}.
Understanding the mechanisms governing triboelectric charging in turbulent particle-laden flows is therefore essential for the design of safe and efficient industrial processes. 

Experiments have shown that flow velocity, i.e., Reynolds number, is one of the most important and controllable parameters affecting particle charging and, hence, the hazard of dust explosions \citep{Schwindt2017,Fath2013}.
However, most numerical investigations using Direct Numerical Simulations (DNS) have focused, to reduce computational costs, on the low-turbulent regime with a friction Reynolds number of \(\Rey_\tau = 180\) or lower \citep{Grosshans2017, Ozler2025, Jin17}. Investigations at higher Reynolds numbers (\(\Rey_\tau \approx 550\)) showed significant influence of the particle concentration profiles. Nonetheless, the Reynolds number was fixed. As the particle charge was prescribed and constant, its influence on charging was not investigated \citep{Zhang2023, Cui2024}.

The few available numerical studies regarding the effect of Reynolds number on triboelectric charging have reported contradictory results.
\citet{Tanoue2001} observed in Reynolds-averaged Navier-Stokes (RANS) simulations a decrease in charge with increasing Reynolds number.
Contrary, \citet{Watano2006} found no significant dependence in Discrete Element Modelling (DEM) simulations, and the Large-Eddy Simulations (LES) by \citet{Grosshans2016b} showed an increase in charge with increasing Reynolds number.
Clarifying these discrepancies requires a systematic investigation of how different Reynolds numbers affect the particle distribution and charging rate.

Advances in computational resources over the past decades have made it feasible to simulate triboelectric charging in increasingly complex systems. 
Coupled computational fluid dynamics--discrete element method (CFD--DEM) approaches have
become a common tool to investigate triboelectric charging at the fundamental level, as they allow for the coupling of the carrier flow dynamics with the motion of individual particles, including electrostatic interactions. Since particles acquire most of their charge
during collisions and collision dynamics are strongly influenced by the
surrounding flow, accurate modelling of the fluid phase is crucial.

Early RANS simulations~\citep{Tanoue2001} model all turbulent motions.
More recent LES resolve the large-scale turbulent structures in space and time
\citep{Grosshans2016c, Korevaar2014,Grosshans2016b}.
In these LES~\citep{Grosshans2016b}, the increase in the air velocity led to intensified turbulent fluctuations and thereby increased wall-normal particle velocities, impact velocities, and effective contact areas during collisions.
However, these LES did not resolve the viscous sub- and buffer layer, although this region exhibits the highest turbulence intensities and strongest velocity gradients, which can strongly influence charging. Furthermore,
particle--particle collisions were neglected~\citep{Grosshans2016b}, even though they may
significantly affect the particle distribution, even in dilute regimes
\citep{Rupp2023}. 
Resolving all turbulent scales, DNS revealed that triboelectric charging is strongly affected by small-scale motions~\citep{Grosshans2017, Jantac2024}.
Reliable prediction of charging in particle-laden turbulent flows, therefore, requires DNS or LES with a resolved viscous sublayer.
However, the high computational cost of DNS limits such studies to low Reynolds numbers and simple geometries, such as channels, ducts, or straight pipes.

From an industrial perspective, flows at high Reynolds numbers are of particular relevance. Pneumatic conveying systems operating in the dilute regime typically employ air velocities in the range of 10~m s\(^{-1}\) to 30~m s\(^{-1}\), which, for a pipe diameter of 0.1~m, correspond to Reynolds
numbers of order \(\Rey=\textit{O}(10^5)\) \citep{Kli18,Jones1997}.
This range underscores the
need to investigate triboelectric charging at higher
Reynolds numbers than those addressed in previous DNS, as both mixing processes and
turbulence characteristics undergo pronounced changes with increasing
Reynolds number \citep{Lee2015}. 

Also, complex geometries are particularly relevant for industrial applications, as most conveying systems comprise pipe bends, contractions, expansions, and junctions.
These features can strongly influence charging behaviour due to increased impact velocities and contact areas with the walls. 
For example, \citet{Yan2021} showed that electrostatic forces enhance particle
accumulation in the near-wall region of bends, which may increase particle-wall collision frequency and thus charge transfer.

To isolate and quantify the influence of Reynolds number, investigations at different fixed
Reynolds numbers are essential. 
In this context, fully developed conditions of both the gas and particle phases are favoured, as they
eliminate spatial transients and allow changes in charging behaviour to
be attributed unambiguously to Reynolds-number effects. Numerical
simulations offer a clear advantage in this regard. In experimental
configurations, it is essentially impossible to establish a fully
developed state of both phases without simultaneously inducing particle
charging.
Therefore, reproducibility in experiments is challenging because it is difficult to control the electrostatic boundary conditions \citep{Lim2006}.
Moreover, achieving such conditions would require very long ducts, preferably oriented vertically to minimise gravitational segregation within the cross-section, rendering systematic parametric studies impractical. 
In contrast, numerical simulations can employ periodic boundary conditions to fully developed the flow within a compact computational domain.

To date, the only open-source code for simulating triboelectric particle charging \citep{pafiX} is limited to simple geometries, namely channels and rectangular ducts, and lacks turbulence models.
To overcome these limitations, we introduce in this paper a new open-source solver, triboFoam \citep{Wilms2026}.
This new solver is built upon OpenFOAM \citep{Weller1998} and, thus, takes advantage of the framework's capabilities to model turbulence and complex geometries.
In the present work, we apply LES with a fully resolved viscous sublayer to simulate friction Reynolds numbers up to \(\Rey_\tau = 550\), ensuring sufficient numerical resolution. 
In addition, different particle diameters are investigated to assess the influence of particle inertia in interaction with turbulent structures.

The remainder of the paper is organised as follows.
Section~\ref{sec-MathModeling} presents the governing equations of the
fluid and particle phase, including the electrostatic and
charging models. Section~\ref{sec-NumericalSetup} describes the
numerical setup of the DNS and LES simulations, followed by Section~\ref{sec-Results} presenting and discussing the results.
Finally, conclusions are drawn in Section~\ref{sec-Conclusion}.

\section{Mathematical modeling}\label{sec-MathModeling}

The system is modelled using an Eulerian framework for the fluid phase
and a Lagrangian framework to capture the particle motion.
In this approach, the flow is considered to be dilute, meaning that the volume occupied by the particles is very small compared to the fluid volume, which allows us
to consider the particles as point quantities.

\subsection{Fluid phase}\label{fluid-phase}

The fluid phase is described by the incompressible Navier-Stokes equations with constant diffusivities,
\begin{subequations}
\begin{equation}
\nabla \cdot \boldsymbol{u} = 0
\end{equation}
and
\begin{equation}\phantomsection\label{eq-ns}{
		\frac{\partial \boldsymbol{u}}{\partial t} + \left( \boldsymbol{u}\cdot \nabla \right) \boldsymbol{u} =  -\frac{1}{\rho}\nabla p + \nu \nabla^2 \boldsymbol{u} + \boldsymbol{F}_s + \boldsymbol{F}_f.
}\end{equation}
\end{subequations}
In these equations, \(\boldsymbol{u} = (u, v, w)\) represents the fluid
velocity vector, \(p\) the pressure, \(t\) the time and the fluid
density and kinematic viscosity is described by \(\rho\) and \(\nu\),
respectively. 
The source term \(\boldsymbol{F}_s\) considers the momentum
transfer from the particles to the fluid by summing up all
aerodynamic forces over a control volume. The last source term,
\(\boldsymbol{F}_f\), describes a general force acting on the entire fluid,
which is, e.g., in the case of a cyclic channel flow, a pressure gradient to
propel the flow.

The governing equations of the LES are comparable to those of the DNS. The velocity field of the continuous phase can be decomposed into a filtered component (\(\overline{\boldsymbol{u}}\)) and a sub-filter (or sub-grid scale, SGS) component. 
The filtered Navier-Stokes equations read,
\begin{subequations}
\begin{equation}
\nabla \cdot \overline{\boldsymbol{u}} = 0
\end{equation}
\begin{equation}
\frac{\partial \overline{\boldsymbol{u}}}{\partial t} + \left( \overline{\boldsymbol{u}}\cdot \nabla \right) \overline{\boldsymbol{u}} =  -\frac{1}{\rho}\nabla \overline{p} + \nu \nabla^2 \overline{\boldsymbol{u}} - \nabla \cdot \bm{\tau}_\text{SGS} + \boldsymbol{F}_s
\end{equation}
\end{subequations}
with the SGS stress tensor \(\bm{\tau}_\text{SGS}\). The equations
of motion are closed by a model for the SGS stress tensor. Common models
rely on the eddy-viscosity approach, where the SGS tensor is expressed as
\begin{equation}\phantomsection\label{eq-nu-sgs}{
		\bm{\tau}_{\text{SGS}} - \frac{2}{3} k_{\text{SGS}} \mathsfbi{I} = - 2 \nu_{\mathrm{SGS}} \overline{\mathsfbi{S}}
}\end{equation}
with the sub-grid scale eddy viscosity \(\nu_\text{SGS}\), the sub-grid scale turbulent
kinetic energy \(k_{\text{SGS}}\) the identity tensor
\(\mathsfbi{I}\), and the strain rate tensor \(\mathsfbi{S}\), which is
defined as
\begin{equation}
\mathsfbi{S}
= \frac{1}{2}
\left(
\nabla \boldsymbol{u}
+
(\nabla \boldsymbol{u})^{T}
\right).
\end{equation}

To close equation~(\ref{eq-nu-sgs}), OpenFOAM provides several subgrid-scale models, namely the \citet{Smagorinsky1963} model, the \(k\)-equation model \citep{Yoshizawa1986},
the dynamic \(k\)-equation model \citep{Kim1995}, and the Wall Adapting Local Eddy-viscosity (WALE) model \citep{Nicoud1999}. 
Section~\ref{sec-apx-les-models} presents a comparison of these four LES models and shows that, in terms of both velocity profiles and particle distributions, the WALE model matches the DNS results the closest.
Consequently, it is adopted as the default SGS model for all subsequent simulations. 
The WALE model computes the subgrid viscosity proportional to the cube of the wall-normal distance and allows it to vanish in the vicinity of solid walls. The SGS viscosity is given by
\begin{equation}
\nu_\text{SGS}
= \left(c_w \Delta\right)^2
\frac{\left( \mathsfbi{S}^d : \mathsfbi{S}^d \right)^{3/2}}
{\left( \overline{\mathsfbi{S}} : \overline{\mathsfbi{S}} \right)^{5/2}
	+ \left( \mathsfbi{S}^d : \mathsfbi{S}^d \right)^{5/4}}
\end{equation}
with the model constant \(c_w = 0.325\), the filter width \(\Delta = \mathcal{V}^{1/3}\), with the cell volume \(\mathcal{V}\), and the traceless symmetric part
of the square of the velocity gradient tensor
\begin{equation}
\mathsfbi{S}^d
= \frac{1}{2}
\left(
\left(\nabla \overline{\boldsymbol{u}}\right)^2
+
\left(\left(\nabla \overline{\boldsymbol{u}}\right)^2\right)^{T}
\right)
- \frac{1}{3}
\operatorname{tr}\!\left(\left(\nabla \overline{\boldsymbol{u}}\right)^2\right)
\mathsfbi{I}.
\end{equation}

\subsection{Particle phase}\label{particle-phase}

The particle phase is assumed to be an ensemble of multiple rigid and
spherical particles made out of the same material with the same density
\(\rho_p\). The motion of the particles can be described by Newton's
law, \begin{equation}\phantomsection\label{eq-particle-motion}{
		\frac{\text{d} \boldsymbol{u}_p}{\text{d}t} = \boldsymbol{f}_{\text{ad}} + \boldsymbol{f}_{\text{coll}} + \boldsymbol{f}_{g} + \boldsymbol{f}_{\text{el}}
}\end{equation}
with the particle velocity vector \(\boldsymbol{u}_p\) and the specific forces \(\boldsymbol{f}_{\text{ad}}\), \(\boldsymbol{f}_{\text{coll}}\), \(\boldsymbol{f}_{g}\), and \(\boldsymbol{f}_{\text{el}}\) due to aerodynamics, collisions, gravity, and the electric field, respectively.

We considered in our simulations only the drag force, as it has the highest
contribution across all aerodynamic forces.
It is computed by \citep{Crowe2011} 
\begin{equation}
\boldsymbol{f}_\mathrm{ad} = \frac{3}{4}\frac{\rho\,\nu\,C_D\,Re_p}{\rho_p \, d_p^2}
\end{equation}
with the particle Reynolds number \(\Rey_p = \left(|\boldsymbol{u}_\mathrm{rel}|\,d_p\right)/\nu\), the particle velocity relative to the fluid velocity \(\boldsymbol{u}_\mathrm{rel} = \boldsymbol{u}_p - \boldsymbol{u}\), the particle
density \(\rho_p\), and the particle diameter \(d_p\). 
The particle drag coefficient (\(C_D\)) is a function of \(\Rey_p\), computed by \citep{Schiller1933} 
\begin{equation}
\mathrm{C}_\mathrm{D} = \begin{cases}
	\frac{24}{\mathrm{Re}_p}
	\left(1 + \frac{1}{6}\mathrm{Re}_p^{2/3} \right) & \text{for }Re_p\le 1000  \\
	0.424 & \text{for }Re_p > 1000.
\end{cases}
\end{equation}

In the following, quantities with subscript \(w\) refer to the wall, with subscript \(i\) to a particle contacting a wall, and with subscript \({ij}\) to a particle contacting another particle. 
The subscript \({ji}\) is used accordingly to indicate the second particle
in a particle-particle collision. To shorten the notation, the subscript
\(p\) indicates that the particle phase is partially omitted in the
following equations. The collision forces can be distinguished into
binary particle-particle collisions and particle-wall collisions. In
both cases, a hard-sphere model with fully elastic collisions is
considered. In particle-wall collisions, the wall-normal velocity
component flips its sign while the wall-tangential components remain
unchanged \citep{Grosshans2017}. In particle-particle collisions, the velocity vectors are updated according to Hertzian theory, where we consider monodispersed particles \(r_{ij} = r_{ji}\).

The force due to gravitational acceleration \(\boldsymbol{g}\) is based on
the density ratio of the fluid to the particle and reads as \(\boldsymbol{f}_{g} = \left(1 - \left(\rho/\rho_p\right) \right) \boldsymbol{g}.
\)

The electrostatic force is considered with a hybrid scheme which combines Gauss' and Coulomb's law \citep{Grosshans2017b}. 
Coulomb law is assumed to have no numerical inaccuracy, as it is mesh
independent; however, the computational effort scales with the number of
particles
\(N \left(N - 1 \right) / 2 \sim \textit{O}\left(N^2\right)\), as the force has to be computed for every particle pair. 
The numerical accuracy of Gauss law strongly depends on the mesh resolution,
but only depends linearly on the number of particles \(N\). Hence, the
hybrid approach combines the advantages of both methods by considering
Coulomb law for nearby particles (within the same cell) and the effect
of all other particles in the far field by Gauss law. The force based on
Coulomb law for particle \({ij}\) can be described as 
\begin{equation}
\boldsymbol{f}_{\text{el}, \text{Coulomb},ij} = \frac{q_{ij} q_{ji}}{4 \pi \varepsilon} \frac{\hat{\boldsymbol{s}}_{12}}{|\boldsymbol{s}_{ij}-\boldsymbol{s}_{ji}|^2}
\end{equation}
where \(\boldsymbol{s}\) describes the location of the particle centre,
\(\hat{\boldsymbol{s}}_{12}\) a unit vector from particle \({ij}\) to
\({ji}\), \(q\) the electric charge of the individual particles, and
\(\varepsilon\) the electric permittivity of air, which is assumed to be equal 
to the permittivity of the vacuum, \(\varepsilon_0\). The force for the other
particle is, according to Newton's third law,
\(\boldsymbol{f}_{\text{el},\text{Coulomb},ji} = -\boldsymbol{f}_{\text{el},\text{Coulomb},ij}\).
The calculation via Gauss's law utilised the electric field created by
the charged particles: 
\begin{equation}
\boldsymbol{f}_{\text{el},\text{Gauss}} = \frac{q \boldsymbol{E}}{m_p}
\end{equation}
with the particle mass \(m_p\), in case of spherical particles
\(m_p = \frac{1}{6} \pi d_p^3 \rho_p\). The electric field
strength, \(\boldsymbol{E}\), equals the negative gradient of the electric
potential, \(\phi\), i.e., \(\boldsymbol{E} = - \nabla \phi\).
	
The relation between the electric potential and the electric charge
density \(\rho_{\text{el}}\) is given by the Poisson equation 
\(\nabla^2\phi = -\left(\rho_{\text{el}}/\varepsilon\right)\).
The integration of \(\rho_{\text{el}}\) over a control volume \(\mathcal{V}\),
which contains \(n\) particles, is equivalent to the sum of the charges
of the \(n\) particles. In this case, the control volume \(\mathcal{V}\) is represented by a
grid cell. 

In the following, various levels of coupling between the fluid and
particle phases are considered. In 1-way coupling, particle motion is
influenced solely by the aerodynamic force, \(\boldsymbol{f}_{\text{ad}}\), and
the gravitational force, \(\boldsymbol{f}_{g}\). In 2-way coupling, the
particle momentum is fed back to the fluid through the term
\(\boldsymbol{F}_s\) in equation~\ref{eq-ns}. In 4-way coupling, particles
also interact with one another via the particle--particle collision
force, \(\boldsymbol{f}_{\text{coll}}\), in equation~(\ref{eq-particle-motion}), as
well as through the electric force term, \(\boldsymbol{f}_{\text{el}}\), in
equation~(\ref{eq-particle-motion}).

\subsection{Charging models}\label{charging-models}

The effect of triboelectric charging has been implemented in various models. We have implemented in triboFoam the conventional condenser model \citep{Soo1971}, a simple random charging model \citep{Ozler2025}, a model which transfers a constant charge, and the recently published Stochastic Scaling Model (SSM) \citep{Grosshans2025}.

\subsubsection{Condenser model}\label{condenser-model}

The condenser model computes the charge transfer in a particle-particle
collision by
\begin{equation}
\Delta q_{ij} = - \Delta q_{ji} = \frac{C_{ij} C_{ji}}{C_{ij} + C_{ji}}\left( \frac{q_{ji}}{C_{ji}} - \frac{q_{ij}}{C_{ij}}\right) \left( 1 - e^{-\Delta t_{12} / T_{12}} \right),
\end{equation}
where \(\Delta t_{12}\) is the contact time during particle-particle
collision, \(T_{12}\) is the charge relaxation time, and \(C_{ij}\)
respectively \(C_{ji}\) are the electric capacities of both particles.
The electric capacity of particle \({ij}\) is computed as \(
C_{ij} = 4 \pi \varepsilon d_{ij}/2
\) while \(C_{ji}\) is determined accordingly. Further, the charge
relaxation time \(T_{12}\) is expressed by 
\begin{equation}
T_{12} = \frac{C_{ij} C_{ji}}{C_{ij} + C_{ji}} \frac{d_{ij} + d_{ji}}{2A_{12}} \varphi_p,
\end{equation}
where the resistivity of the particle is denoted by \(\varphi_p\). The
calculation of the contact surface \(A_{12}\) is performed in accordance
with the elastic theory of Hertz as
\begin{equation}
A_{12} = \frac{\pi d_{ij} d_{ji}}{2\left(d_{ij} + d_{ji}\right)} \alpha_{12}
\end{equation}
with
\begin{equation}
\alpha_{12} =  \frac{d_{ij} d_{ji}}{2} \left( \frac{5}{8} \pi \rho_p \left(1 + k_e\right) |\boldsymbol{u}_{12}|^2 \frac{\sqrt{d_{ij} + d_{ji}}}{d^3_{ij} + d^3_{ji}} \frac{1-\mu_p}{E_p} \right)^{2/5}.
\end{equation}
In this expression, \(\boldsymbol{u}_{12}\)\hspace{0pt} denotes the
relative velocity between the two colliding particles, defined as
\(\boldsymbol{u}_{12} = \boldsymbol{u}_{ji} - \boldsymbol{u}_{ij}\)\hspace{0pt}.
The parameters \(\mu_p\) and \(E_p\)\hspace{0pt} represent,
respectively, the Poisson ratio and Young's modulus of the particle
material. The contact duration, \(\Delta t_{12}\), is obtained from
Hertzian contact theory as \(
\Delta t_{12} = \left(2.94 \alpha_{12}\right)/\left(|\boldsymbol{u}_{12}|\right).
\)

The computation of charge transfer during particle--wall collisions
follows the model proposed by \citet{John1980}, obtained from the
particle--particle charge exchange formulation of \citet{Soo1971}, in the limit where the diameter of one particle tends to infinity. The total charge exchange, \(\Delta q_w\), is expressed as the sum of two distinct
contributions,
\begin{equation}
\Delta q_w = \Delta q_c + \Delta q_t.
\end{equation}
Here, \(\Delta q_c\)\hspace{0pt} denotes the dynamic charge transfer due to the contact potential and \(\Delta q_t\)\hspace{0pt} the transfer of the particle's pre-existing charge. The
contact area between the particle and the channel wall is assumed to be
small compared to the particle surface area.
Consequently, the dynamic
charge transfer is modelled analogously to the charging of a
parallel-plate capacitor. In this framework, the particle--wall charge
exchange, \(\Delta q_c\)\hspace{0pt}, is
\begin{equation}\phantomsection\label{eq-particle-wall-exchange}{
		\Delta q_c = -C U_c \left( 1 - e^{-\Delta t_w / T_w}\right)
}\end{equation}
where \(C\) is the electrical capacitance, \(U_c\)\hspace{0pt} is the
particle--wall contact potential, \(\Delta t_w\)\hspace{0pt} is the
duration of the particle--wall collision, and \(T_w\) is the charge
relaxation time. The capacitance of a parallel--plate capacitor is given
by
\begin{equation}\phantomsection\label{eq-capacity-capacitor}{
		C = \frac{\varepsilon A_{iw}}{h}
}\end{equation}
where \(h\) denotes the separation between the capacitor plates and
\(A_{iw}\) is the plate area. In the present context, \(h\) corresponds
to the effective particle--wall separation during impact, while
\(A_{\text{iw}}\) represents the contact area between the particle and
the wall. The latter is estimated from Hertzian elastic theory (see
\citet{John1980}) by
\begin{equation}\phantomsection\label{eq-charge-contact-area}{
		A_{iw} = \pi \frac{d_p}{2} \alpha_{iw}
}\end{equation}
with
\begin{equation}\phantomsection\label{eq-charge-contact-area-alpha2}{
		\alpha_{iw} = \frac{d_p}{2} \left( \frac{5}{8} \pi \rho_p \left( 1 + k_e \right) |\boldsymbol{u}_i|^2 \left( \frac{1 - \mu_p^2}{E_p} + \frac{1 - \mu_w^2}{E_w} \right) \right)^{2/5} \, .
}\end{equation}
The parameters \(\mu_w\) and \(E_w\) represent the Poisson ratio and
Young's modulus, respectively, of the wall material. Following the
arguments of \citet{John1980}, the plate separation distance \(h\) in
equation~(\ref{eq-capacity-capacitor}) is taken to be of the order of the
range of repulsive molecular forces arising from surface irregularities.
Hertzian contact theory is also applied to estimate the collision
duration \(\Delta t_w\)\hspace{0pt} appearing in
equation~(\ref{eq-particle-wall-exchange}). According to this theory,
\(\Delta t_w\)\hspace{0pt} is given by
\begin{equation}
\Delta t_w = \frac{2.94}{|\boldsymbol{u}_i|} \alpha_{iw} \, .
\end{equation}
Furthermore, the charge relaxation time \(T_w\) appearing in
equation~(\ref{eq-particle-wall-exchange}) is determined by \(T_w = \varepsilon_w \varepsilon_0 \varphi_p\).
In this expression, \(\varepsilon_w\) denotes the relative permittivity of
the plates, and \(\varphi_p\) is the electrical resistivity of the
particle.

The particle pre-charge, \(q\), refers to the charge present
prior to its collision with the wall, and is assumed to be uniformly
distributed over its surface. The corresponding pre-charge transfer
across the contact surface, \(\Delta q_t\)\hspace{0pt}, is therefore
given by

\begin{equation}\phantomsection\label{eq-precharge-surface}{
		\Delta q_t = - \frac{1}{4}\alpha_{iw} q.
}\end{equation}

The assumption of a uniformly distributed pre-charge does not generally
hold for non-conducting particle surfaces \citep{Grosshans2016}.
For the cases considered in the present study, however, the charge accumulated by each particle remains much smaller than its equilibrium value. As a
result, the contribution of \(\Delta q_t\)\hspace{0pt} to the total
charge transfer is negligible compared with that of
\(\Delta q_c\)\hspace{0pt}, and the use of 
equation~(\ref{eq-precharge-surface}) is sufficiently accurate for the
purposes of the present analysis.

\subsubsection{Stochastic scaling model}\label{stochastic-scaling-model}

Contrary to the condenser model, the new SSM \citep{Grosshans2025} can predict three experimentally observed charging patterns:
(a) scattered particle-wall impact charge \citep{Grosjean2023}, (b) size-dependent particle-particle
(bipolar) charging \citep{Forward2009}, and (c) charge reversal after some wall impacts \citep{Shaw1928}.
The SSM unifies the charging of particles of the same material (particle-particle) and of different materials (particle-wall) in a single framework. This framework is based on a stochastic closure
involving the mean, variance, skewness, and minimum impact charge obtained from a single reference experiment.

Similar to the mosaic \citep{Apo10,Grosj23} and surface state models~\citep{Low86a,Low86b,Lacks08,Lacks07}, the contact area between
the particle and another surface consists of charge-exchange sites. When
particles collide with a wall, each of the \(N_w\) active charging sites
on the wall contributes a charge \(\epsilon_w\) to the particle. Hence,
the total transfer from the wall to the particle equals to \(
\Delta q_w = \epsilon_w N_w \sim F\left( \mu_w \right),
\) where \(\mu_w\) corresponds to the expected value of \(\Delta q_w\).
In the case of a conductive and grounded wall, the charging sites regenerate immediately upon contact. This maintains a consistent number
of active sites both before and after impact.

On the other hand, according to SSM, charge transfer from an insulative particle surface is probabilistic. Each of the \(N_{ij}\) active
charging sites that make up the contact area of particle \({ij}\) can
either transfer a charge \(\epsilon_{ij}\) to the opposing surface or
not. This stochastic process follows a binomial distribution, which is
approximated by a skewed normal distribution characterised by a mean
\(\mu_{ij}\), a standard deviation \(\sigma_{ij}\) and a skewness
\(\gamma_{ij}\) and can be expressed as 
\begin{equation}
\Delta q_1 = \epsilon_p \sum _{n=1} ^{N_{ij}} \theta_n \sim G\left( \mu_{ij}, \sigma_{ij}, \gamma_{ij} \right),
\end{equation}
with the Bernoulli random variable \(\theta_n\) of equal probability of
0 or 1.

To obtain the charge transfer during an individual collision, the
statistical parameters \(\mu_w\), \(\mu_{ij}\), \(\sigma_{ij}\), and
\(\gamma_{ij}\) are scaled from a reference experiment with known
parameters \(\mu_0\), \(\sigma_0\), \(\gamma_0\), \(N_{w0}\), and
\(N_0\) to the actual number of active charging sites \(N_w\) and
\(N_{ij}\)by the relations
\begin{subequations}
\begin{equation}
\mu_w = \Delta q_{0, \text{min}} \frac{N_w}{N_{w0}}
\end{equation}
\begin{equation}
\mu_i = \left( \Delta q_{0, \text{min}} - \mu_0 \right)\frac{N_i}{N_0}
\end{equation}
\begin{equation}
\sigma_i = \sigma_0 \sqrt{\frac{N_i}{N_0}}
\end{equation}
\begin{equation}
\gamma_i = -\gamma_0 \frac{N_0}{N_i}
\end{equation}
with
\begin{equation}
\frac{N_w}{N_{w0}} = \frac{A}{A_0}\frac{\chi}{\chi_0} \qquad \text{and} \qquad \frac{N_i}{N_0} = \frac{A}{A_0}\frac{\chi}{\chi_0}\frac{c}{c_0} \, .
\end{equation}
\end{subequations}
The contact area \(A\) is estimated from Hertzian contact theory as
described above.

The activity ratio \(\chi/\chi_0\), the ratio of active to total charging sites, is obtained from
\begin{subeqnarray}
	\frac{\chi}{\chi_0} = \frac{\chi_w}{\chi_0} &=& 1 - \text{sgn}\left(\Delta q_{0, \text{min}} \right) \frac{4q_i}{C_\text{sat}d_i^2} \\
	\frac{\chi}{\chi_0} = \frac{\chi_i}{\chi_0} &=& 1 + \text{sgn}\left(\Delta q_{0, \text{min}} - \mu_0\right) \frac{4q_i}{C_\text{sat}d_i^2} \\
	\frac{\chi}{\chi_0} = \frac{\chi_{ij}}{\chi_0} &=& 1 - \text{sgn}\left(\Delta q_{0, \text{min}} - \mu_0\right) \frac{8}{C_\text{sat}} \left( \frac{q_j}{d_j^2} - \frac{q_i}{d_i^2}\right)
\end{subeqnarray}
with \(0 \le \chi/\chi_0 \le 1\), where \(C_\text{sat}\) denotes a
fitting parameter. Numerical simulations for air at atmospheric pressure
and unit permittivity yielded
\(C_\text{sat} = 500\,\si{\micro\coulomb}\text{\,m}^{-2}\) \citep{Matsuyama2003}.

The charging site density \(c/c_0\) (\(0 \le c/c_0 \le 1\)) is a
function of the number of collisions \(M\) and is computed for the
\(m\)th contact as
\begin{equation}
\frac{c}{c_0}\left(t_{M-1} \le t \le t_M\right) = 1 - \sum_{m=1}^{M-1} \frac{c_{m-1}}{c_0}\frac{A_m}{A_p}\frac{\chi_m}{\chi_0}\frac{q_{i,m}}{q_{i,\text{max}}}
\end{equation}
where \(A_m/A_p\) is the ratio of the contact area to the particle surface area.

The reference experiment for determining the parameters with the subscript
\(0\) consists of the normal impact of a single particle onto a fixed
plate. Particles of diameter \(d_0\) are released onto the target, and the
wall-normal impact velocity \(u_{n,0}\) is obtained from high-speed
imaging. The associated charge transfer is quantified following the
procedure of \citet{Matsuyama2003}. Repeated realisations of the
experiment yield statistical distributions from which \(\mu_0\),
\(\sigma_0\), \(\gamma_0\), and \(\Delta q_{0, \text{min}}\) are
evaluated. Further details of the model formulation and experimental
configuration are provided by \citet{Grosshans2025}.

\section{Numerical setup}\label{sec-NumericalSetup}

Two different numerical setups are utilised in this work.
The first validates the triboFoam solver using DNS.
The second uses LES to study the effects of higher Reynolds numbers on particle mechanics and charging. 
The subsequent sections describe both setups.

\subsection{\texorpdfstring{DNS setup for the validation of	triboFoam}{DNS setup for the validation of triboFoam}}\label{sec-ValidationSetup}

The solver triboFoam is validated against pafiX, because, to the best of our knowledge, pafiX is the only existing open-source solver for triboelectric particle charging.
The motion of the fluid and uncharged particles has been validated against literature data \citep{Sardina2012,Vreman2014}.
Since pafiX is a DNS solver and to avoid interference with a turbulence model, triboFoam validation is based on DNS of turbulent channel flows, first for uncharged and then for charged particles.

The channel flows are fully-developed at a fixed friction Reynolds number of \(\Rey_\tau = u_\tau \delta / \nu = 180\) with the friction velocity \(u_\tau = \sqrt{\tau_w / \rho}\), where \(\tau_w\) is the wall shear stress, and the half channel height \(\delta\). 
In the following, quantities indicated with the superscript \(^+\) are dimensionless with
\(u_\tau\) to obtain values in wall units. 
Hence, the streamwise velocity \(u^+\) is defined as \(u^+ = u / u_\tau\), the wall-normal distance as \(y^+ = \left( y u_\tau\right) / \nu\), and the time as \(t^+ = \left( t u_\tau^2 \right) / \nu\).

The size of the computational domain is \(12\delta \times 2\delta \times 4\delta\) in streamwise (\(x\)), wall-normal (\(y\)), and spanwise (\(z\)) direction with \(\delta = 0.02\,\)m.
In streamwise and spanwise directions, periodic boundary conditions are applied, while the wall features the no-slip
condition. 
The fluid is driven by a constant pressure gradient, and its properties are set to be equivalent to those of air.
The numerical values of all material parameters in the simulations are summarised in table~\ref{tbl-material-properties}.
The spatial resolution in stream- and spanwise direction amounts to \(\Delta x^+ \approx 8.44\) and \(\Delta z^+ \approx 5\), respectively. 
In wall-normal direction, the size of the first cell at the wall equals \(\Delta y^+ \approx 0.036\) and in the channel centre \(\Delta y^+ \approx 4\).
All derivatives of the Navier-Stokes and the electrostatic equations are discretised second order in time and space using finite volumes.

Three different particle sizes were investigated, \(d_p = 25 \,\unit{\um}, 50 \,\unit{\um}, 100 \,\unit{\um}\) while keeping the material density (table~\ref{tbl-material-properties}) and the particle number density \((C = 10^8\, \text{m}^{-3})\) constant.
In this case, the gravity vector is set to zero. This setup was chosen as it represents realistic conditions even though the particle volume fraction \(\omega\) rises with increasing \(d_p\). The values set for the particles correspond to the material polymethylmethacrylate (PMMA), which is a typical plastic transported by pneumatic conveying.

\begin{table}
	\begin{center}
		\def~{\hphantom{0}}
		\begin{tabular}{lll}

	Parameter & Symbol & Value \\[3pt]
	Poisson's ratio, particle & \(\mu_p\) & \(0.4\) \\
	Young's modulus, particle & \(E_p\) &
	\(10^8\, \text{kg}\,\text{s}^{-2}\,\text{m}^{-1}\) \\
	Resistivity, particle & \(\varphi_p\) &
	\(10^{13}\, \varOmega\,\text{m}\) \\
	Particle restitution ratio & \(k_e\) & \(0.95\) \\
	Poisson's ratio, wall & \(\mu_w\) & \(0.28\) \\
	Young's modulus, wall & \(E_w\) &
	\(10^{11}\, \text{kg}\,\text{s}^{-2}\,\text{m}^{-1}\) \\
	Relative permittivity, wall & \(\varepsilon_w\) & \(5.0\) \\
	Effective separation & \(h\) & \(10^{-9}\, \text{m}\) \\
	Air permittivity & \(\varepsilon\) &
	\(8.854 \cdot 10^{-12}\, \text{F}\,\text{m}^{-1}\) \\
    Air kinematic viscosity & \(\nu\) & \(1.46\cdot 10^{-5}\, \text{m}^2\,\text{s}^{-1}\) \\
    Air density & \(\rho\) & \(1.2\,\text{kg}\,\text{m}^{-3}\) \\
    Particle density & \(\rho_p\) & \(1150\,\text{kg}\,\text{m}^{-3}\)	\\
  \end{tabular}
 			\caption{\label{tbl-material-properties}Material properties for the simulations  \citep{Grosshans2017}.}
\end{center}
\end{table}

The Stokes number, \(St\), is defined as the ratio between particle response time, \(T_r = \frac{1}{18}\rho_p d_p^2 \rho^{-1} \nu^{-1}\), and a characteristic time scale of the fluid flow, \(T_f = \delta u_c^{-1}\), with the centreline velocity \(u_c\), resulting in
\begin{equation}
St = \frac{1}{18}\frac{\rho_p}{\rho}\frac{d_p^2}{\nu}\frac{u_c}{\delta}.
\end{equation}

Due to different particle sizes, the Stokes number differs per simulation. 
For Stokes numbers much greater than unity, particle trajectories are only weakly influenced by the surrounding flow. Conversely, when the Stokes number is much less than unity, particle motion closely follows the pathlines of the flow.
The Stokes numbers and other dimensional quantities for the different particle sizes are given in table~\ref{tbl-particle-nondimensional}.

\begin{table}
	\begin{center}
		\def~{\hphantom{0}}
    		\begin{tabular}{lrrrrrr}
			& \(\Rey_\tau\) & \(d_p / \text{\textmu} \text{m}\)  & \(d_p^+\) & \(u_c / (\text{m}/\text{s})\) & \(St\) & \(\omega \) \\[3pt]
    \multirow{3}{*}{DNS} & 180 & 25 & 0.22 & 2.50 & 0.28 & 0.82 \\
	& 180 & 50 & 0.45 & 2.50 & 1.14 & 6.54 \\
	& 180 & 100 & 0.90 & 2.50 & 4.56 & 52.36 \\
       & & & & & & \\ 
	\multirow{12}{*}{LES} & 180 & 25 & 0.22 & 2.38 & 0.27 & 0.82 \\
	& 180 & 50 & 0.45 & 2.38 & 1.08 & 6.54 \\
	& 180 & 100 & 0.90 & 2.38 & 4.34 & 52.36 \\
	& 300 & 25 & 0.37 & 4.39 & 0.50 & 0.82 \\
	& 300 & 50 & 0.75 & 4.37 & 1.99 & 6.54 \\
	& 300 & 100 & 1.50 & 4.36 & 7.95 & 52.36 \\
	& 395 & 25 & 0.49 & 5.91 & 0.67 & 0.82 \\
	& 395 & 50 & 0.99 & 5.92 & 2.70 & 6.54 \\
	& 395 & 100 & 1.97 & 5.94 & 10.83 & 52.36 \\
	& 550 & 25 & 0.69 & 8.38 & 0.95 & 0.82 \\
	& 550 & 50 & 1.37 & 8.38 & 3.82 & 6.54 \\
	& 550 & 100 & 2.75 & 8.41 & 15.33 & 52.36 \\
		\end{tabular}
        
	\caption{\label{tbl-particle-nondimensional} Overview of the non-dimensional particle parameters for DNS at
 	\(\Rey_\tau = 180\) and LES from \(\Rey_\tau = 180\) to \(\Rey_\tau = 550\)}
	\end{center}
\end{table}

The simulations on particle charging start with a fully-developed fluid and particle flow field as initial condition.
The particles have an initial charge of \(q = 0\).
In the case of the condenser model, a saturation charge
of \(q_{\text{sat}} = 1.26^{-13}\) C is set which corresponds to about
\(10\,\%\) of the maximum charge a \(d_p = 50\,\unit{\um}\) particle can
hold \citep{Matsuyama2018}. For the SSM, the
reference parameters are chosen as \(r_0 = 2\cdot 10^{-4}\,\text{m}\),
\(u_{n,0} = 0.925\,\text{m}\,\text{s}^{-1}\),
\(\mu_0 = -4.12\cdot 10^{-15}\,\text{C}\),
\(\sigma_0 = 6.49\cdot 10^{-15}\,\text{C}\), \(\gamma_0 = -0.191\), and
\(\Delta q_{0,\text{min}} = -2.19\cdot 10^{-14}\,\text{C}\)
\citep{Grosshans2025}. Inherently, the SSM
considers a size-dependent saturation charge due to its formulation. For
the chosen reference experiment parameters, a \(d_p = 25\,\unit{\um}\)
particle reaches a saturation charge of about \(7.81 \cdot 10^{-15}\)
C, a \(d_p = 50\,\unit{\um}\) particle \(3.13 \cdot 10^{-13}\) C, and a
\(d_p = 100\,\unit{\um}\) particle \(1.25 \cdot 10^{-12}\) C.

The simulations presented after the validation of the SSM employ
the condenser model. Owing to its fully deterministic formulation, this
model allows for a clear attribution of the observed charging behaviour
to the underlying physical mechanisms and thus facilitates a deeper
understanding of the factors contributing to particle charging. One of
the defining features of the SSM is its ability to account for variable
impact charging. Nevertheless, the SSM can be rendered fully
deterministic by setting \(\sigma_0 = \gamma_0 = 0\). Under this
restriction, charging behaviour closely resembling that of the condenser
model can be recovered by choosing
\(\mu_0 = \Delta q_{0,\text{min}} = \mu_{\text{condenser}}\). The
parameter \(\mu_{\text{condenser}}\) depends on the wall-normal impact
velocity \(u_n\) and the particle diameter \(d_p\). Here,
\(u_n = 0.01\,\text{m}\,\text{s}^{-1}\) is selected to match the average
wall-normal particle impact velocity observed in the simulations. For
this configuration, particles with diameters \(d_p = 25\,\unit{\um}\)
require \(\mu_{\text{condenser}} = 2.77 \cdot 10^{-10}\,\text{C}\),
particles with \(d_p = 50\,\unit{\um}\) require
\(\mu_{\text{condenser}} = 2.14 \cdot 10^{-13}\,\text{C}\), and
particles with \(d_p = 100\,\unit{\um}\) require
\(\mu_{\text{condenser}} = 7.71 \cdot 10^{-15}\,\text{C}\).

Section~\ref{sec-apx-condenser-vs-ssm} shows the evolution of the particle
charge as a function of the number of impacts for this parameter
constellation. The results indicate that the charging behaviour
predicted by the two models is nearly identical up to approximately
\(10^4\) collisions. Since such a high number of impacts is never
attained in the simulations considered here, the condenser model and the
deterministic limit of the SSM can be regarded as equivalent under the
present conditions.

\subsection{LES setup for varying Reynolds numbers}\label{sec-LES-setup}

For the LES presented herein, we established the computational grid through a grid-sensitivity study in which the predicted velocity field and particle distribution were compared to results from DNS.
A detailed description of this procedure is provided in Section~\ref{sec-apx-dns-grid-study}.
The study indicated that streamwise and spanwise resolutions of \(\Delta x^+ \approx 9.88\) and \(\Delta z^+ \approx 6.58\), respectively, are sufficient. 
In the wall-normal direction, the first cell adjacent to the wall has a size of \(\Delta y_w^+ \approx 0.47\), while at the channel centre the spacing is \(\Delta y_c^+ \approx 6.25\). This resolution is adequate for all friction Reynolds numbers considered, up to \(\Rey_\tau = 550\).
Moreover, it was observed that with increasing \(\Rey_\tau\), the grid resolution in wall units could be further reduced without loss of accuracy, especially in streamwise direction.
Hence, simulations at \(\Rey_\tau = 550\) were conducted with a streamwise resolution of \(\Delta x^+ \approx 19.75\).

\section{Presentation and discussion of the numerical
	results}\label{sec-Results}

In this section the obtained results are presented and discussed
starting with the validation of the triboFoam solver using 1-way
coupling. In the next step, the effect of 4-way coupling is assessed,
followed by the validation of the LES setup by a comparison to DNS data
at \(\Rey_\tau = 180\). Finally, the effect of increasing Reynolds number
on triboelectric charging is evaluated.

\subsection{\texorpdfstring{Validation of
		triboFoam}{Validation of triboFoam}}\label{validation-of-tribofoam}

\begin{figure}
	
	\begin{minipage}{0.50\linewidth}
		
		\centering{
			
			\includegraphics[keepaspectratio, width=\textwidth]{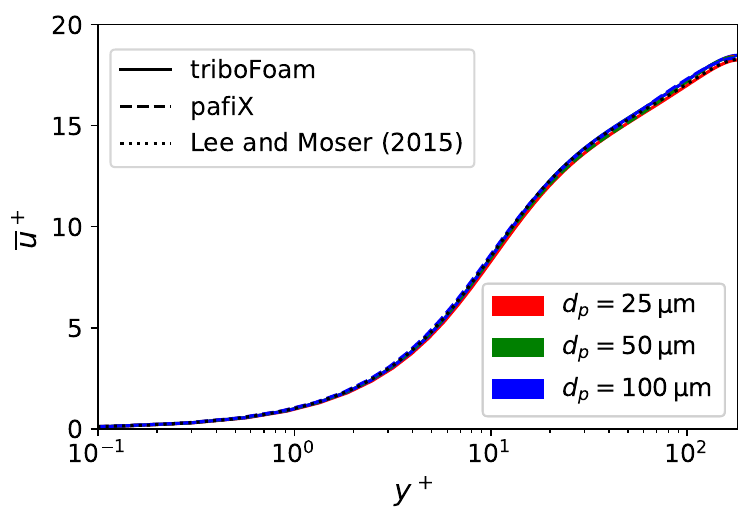}
			
		}
		
		\subcaption{\label{fig-velocity-profile-uncharged}}
		
	\end{minipage}%
	\begin{minipage}{0.50\linewidth}
		
		\centering{
			
			\includegraphics[keepaspectratio, width=\textwidth]{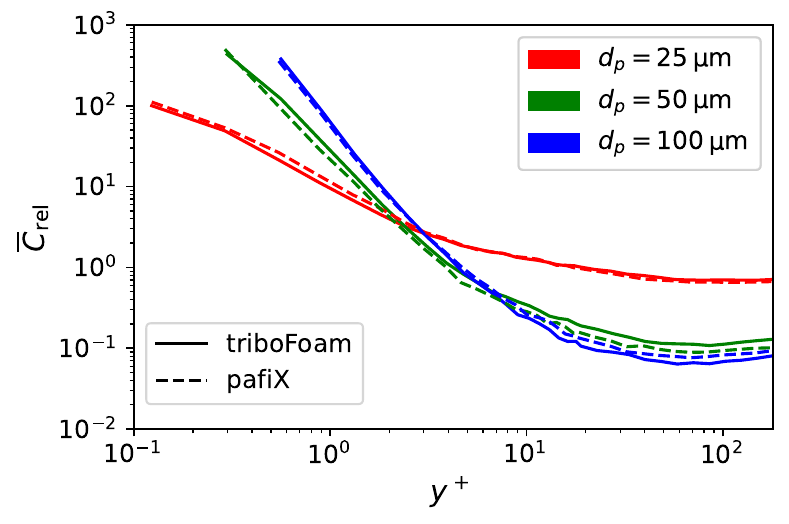}
			
		}
		
		\subcaption{\label{fig-particle-profile-uncharged-1way}}
		
	\end{minipage}%
	
	\caption{\label{fig-velocity-particle-profile-uncharged-1way}(a) Velocity
				\(\overline{u}^+\) and (b) relative particle density \(\overline{C}_\text{rel}\) plotted over wall-normal distance \(y^+\) for triboFoam (solid) and pafiX (dashed) comparing different particle sizes for DNS of \(\Rey_\tau=180\). The particles are 1-way coupled to the fluid. The concentration of the particles is calculated based on the particle centre.}
	
\end{figure}

The first step of the validation considers a fully developed flow of
uncharged particles which are 1-way coupled to the flow, considering
only the drag force, \(\boldsymbol{f}_d\).
Figure~\ref{fig-velocity-profile-uncharged} depicts the mean velocity
profile of the streamwise component over the wall-normal coordinate for triboFoam and pafiX, each with three different particle sizes. The profiles show the typical shape of the universal law of the wall consisting of the viscous sublayer, buffer layer, and logarithmic domain. The comparison to the data of \citet{Lee2015} reveals a maximum
deviation of less than 1 \%. The different particle sizes have no impact
on the profile as the simulations feature only 1-way coupling.

Figure~\ref{fig-particle-profile-uncharged-1way} shows the relative
particle concentration, \(\overline{C}_\text{rel}\), as a function of
wall-normal distance. The relative concentration is defined as

\begin{equation}
\overline{C}_\text{rel} = \frac{\overline{C}}{C_0},
\end{equation}

\noindent where \(\overline{C}\) denotes the temporally averaged particle
concentration at wall-normal location \(y\), and \(C_0\) is the mean
particle concentration across the entire channel. The distribution is
obtained from a histogram whose bins coincide with the cells of the
computational grid, which are identical for both solvers. The results
demonstrate close agreement between the two solvers.
In all simulations, an particle concentration increases in the near-wall region due to turbophoretic drift \citep{Li2021, Ozler2025}, with a pronounced peak one particle radius from the wall, \(y^+ = d_p^+/2\).
Consequently, as the particle diameter increases, the peak shifts
progressively towards the channel centre. With decreasing particle
diameter, the distribution becomes more uniform, which is consistent
with the observations of \citet{Sardina2012}.

\begin{figure}
	
	\begin{minipage}{0.50\linewidth}
		
		\centering{
			
			\includegraphics[keepaspectratio, width=\textwidth]{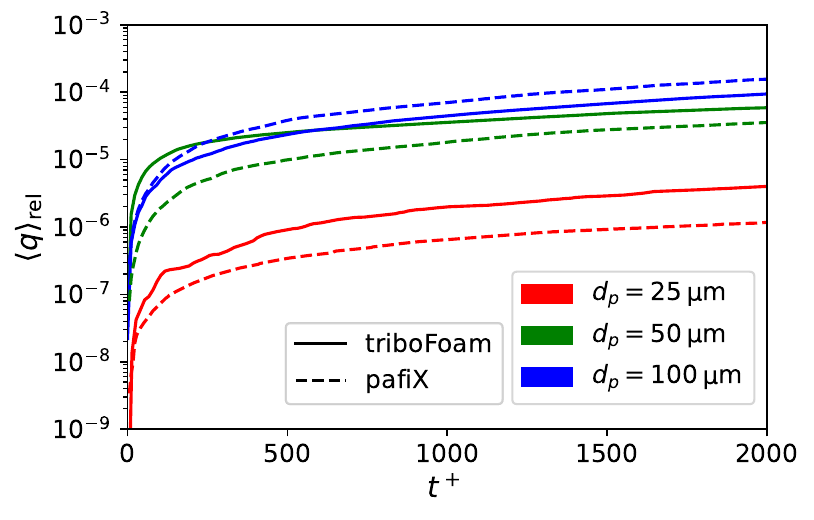}
			
		}
		
		\subcaption{\label{fig-charging-profile-oneway-condenser-ssm-e13-1}Condenser model}
		
	\end{minipage}%
	\begin{minipage}{0.50\linewidth}
		
		\centering{
			
			\includegraphics[keepaspectratio, width=\textwidth]{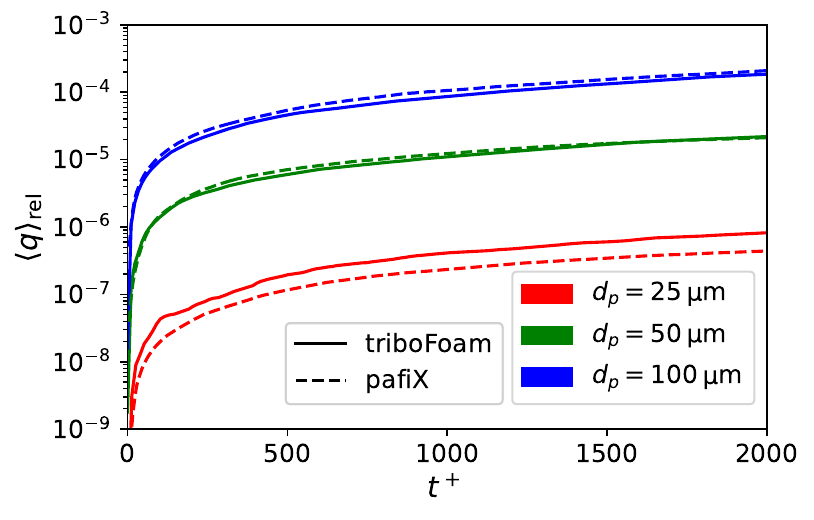}
			
		}
		
		\subcaption{\label{fig-charging-profile-oneway-condenser-ssm-e13-2}SSM}
		
	\end{minipage}%
	
	\caption{\label{fig-charging-profile-oneway-condenser-ssm-e13}Averaged
		relative particle charge \(\langle q \rangle_\text{rel}\) over time for both solvers and different particle sizes for DNS of \(\Rey_\tau=180\) using (a) the condenser model and (b) the SSM.
		The particles are 1-way coupled to the fluid.
        }
	
\end{figure}

In the final step, the charging models are validated by comparing the
temporal evolution of the particle charge. To this end, the average
charge across all particles, \(\langle q \rangle\), is normalised by the
saturation charge, \(q_\text{sat}\), to obtain the relative charge
\(\langle q \rangle_\text{rel} = \langle q \rangle / q_\text{sat}\).
Figure~\ref{fig-charging-profile-oneway-condenser-ssm-e13} compares the
charge build-up for (a) the condenser model and (b) the stochastic
scaling model across both solvers for all three particle sizes. Both
charging models and both solvers exhibit the same overall physical
trend: within the investigated diameter range, larger particles charge
more rapidly than smaller ones. This is attributed to two effects.
First, larger particles experience higher collision velocities, which,
according to Hertzian contact theory, lead to larger contact areas and
thus greater charge transfer per collision (see
equations~(\ref{eq-charge-contact-area}) and
(\ref{eq-charge-contact-area-alpha2})). Second, the collision
frequency increases when the particle diameter rises from
\(d_p = 25\,\unit{\um}\) to \(50\,\unit{\um}\). Increasing the size
further to \(d_p = 100\,\unit{\um}\) results in a slightly reduced
collision frequency, which is consistent with the lower near-wall
particle concentration observed in
figure~\ref{fig-particle-profile-uncharged-1way}.

Although the qualitative behaviour is consistent between the solvers,
quantitative differences remain. For \(d_p = 25\) and
\(d_p = 50\,\unit{\um}\), triboFoam predicts a faster charging
rate than pafiX, whereas the opposite behaviour is observed for
\(100\,\unit{\um}\). The collision frequency and
impact velocity depend on the numerical time-step and grid resolution, and
this dependency is stronger for pafiX. Therefore,
an adequately fine resolution in both space and time was selected and
kept constant for all simulations to ensure a consistent basis for
comparison (\(\Delta t^+ \approx 0.11\)). The discrepancies between the solvers are smaller for the
SSM than for the condenser model. This is explained
by the weaker dependence of the SSM on the contact
area, meaning that differences in impact velocity play a smaller role.
In addition, fluctuations in the charge build-up become more pronounced
with decreasing particle size, particularly in triboFoam. Due to
the smaller Stokes number, the motion of these particles is stronger affected by the turbulent fluctuations of the carrier phase,
which leads to a more intermittent collision behaviour and hence a more
irregular charging process. 

The charge build up in the simulation of \(25\,\unit{\um}\) particles using the condenser model shows at the beginning a relatively high charging rate, which can be attributed to a turbulent sweep pushing many particles towards the wall.
Afterwards, the charging rate decreases to a comparable level of pafiX and fits also into the progression of the other two triboFoam simulations (see also figure~\ref{fig-charging-coupling}). 
Overall, the comparison demonstrates that the triboFoam charging predictions agree well to those of pafiX, while triboFoam is less sensitive to time-step size and grid resolution, indicating a more robust numerical behaviour.
As discussed in the introduction, controlled experimental data for particle charging in fully-developed turbulent flow is unavailable due to the impossibility to control the electrical boundary conditions.

In summary, the turbulent fluid flow, distributions of uncharged particles, and particle charge build-up using the condenser model and SSM simulated with triboFoam has been validated successfully with pafiX and experimental data from the literature.

\subsection{Effect of 2-way and 4-way coupling}\label{effect-of-2-way-and-4-way-coupling}

\begin{figure}
	
	\centering{
	
		\begin{minipage}{0.50\linewidth}
			
			\centering{
			
			\includegraphics[keepaspectratio, width=\textwidth]{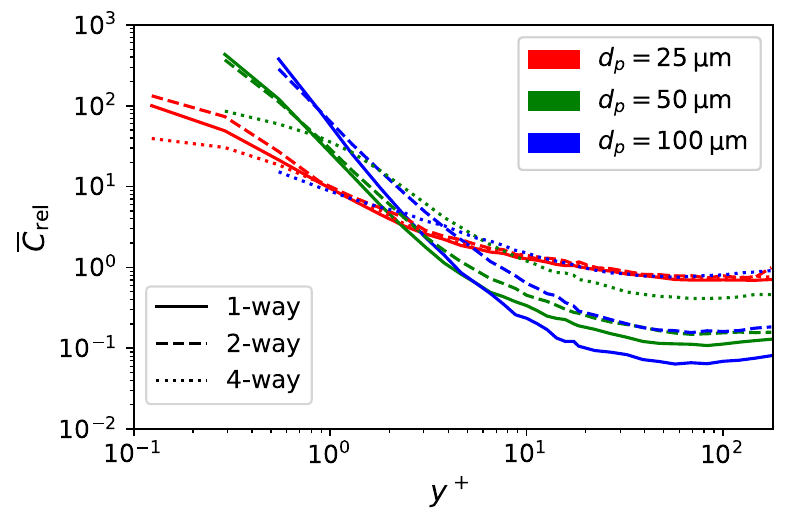}
			
			}
			
		\end{minipage}%
		
	}
	
	\caption{\label{fig-dns-sizes-coupling-particledensity}Relative particle
		density \(\overline{C}_\text{rel}\) plotted over wall-normal coordinate
		\(y^+\) for uncharged particles of different sizes for DNS of \(\Rey_\tau=180\). The coupling between
		the particles and the fluid is varied between 1-way, 2-way, and 4-way.
		The particle concentration profiles are based on the particle
		centre.}

\end{figure}

Up to this point, only 1-way coupling has been considered. In the
following, the effects of 2-way and 4-way coupling are examined.
Figure~\ref{fig-dns-sizes-coupling-particledensity} shows the
wall-normal particle concentration profiles for different particle
sizes. Comparing 1-way and 2-way coupling reveals that the particle
concentration at the location closest to the wall (\(y^+ \approx d_p^+/2\))
decreases for particles with a Stokes number greater than 1 when 2-way
coupling is included. At larger distances from the wall, however, the
concentration increases. This effect
is most pronounced for the largest particles.
Incorporating 4-way coupling further reduces the near-wall concentration, as
particle--particle interactions push particles away from the wall.
This reduction is strongest for the
largest particles, which occupy the highest volume fraction.
Accordingly, the differences between 4-way and 2-way coupling are most
significant for the largest particles, owing to the higher frequency of
particle--particle collisions at increased volume concentration.
These observations are consistent with those of \citet{Rupp2023}.

Moreover, for particles with \(St > 1\), the effect of 4-way versus
2-way coupling is stronger than that of 2-way versus 1-way coupling.
This is partially in accordance with \citet{Rupp2023}, who observed for
particles with \(St = 0.5\) almost no change at \(\Rey_\tau = 300\). The
physical reason for the reduction in particle concentration in the
near-wall region when changing from 1-way to 2-way coupling is
attributed to turbophoresis. 
Particles concentrate in regions of low vorticity at the wall.
2-way coupling reduces these concentration peaks as particles diffuse momentum
from the bulk towards the wall region.
Particle-particle collisions in 4-way coupling even further reduces the near-wall concentration as particles push each other away from the wall.

\begin{figure}
	
	\begin{minipage}{0.50\linewidth}
		
		\centering{
			
			\includegraphics[keepaspectratio, width=\textwidth]{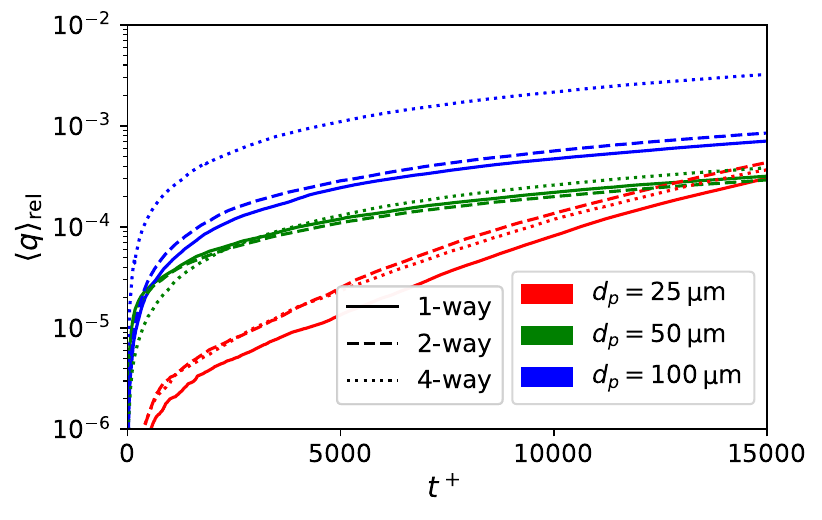}
			
		}
		
		\subcaption{\label{fig-charging-coupling-1}}
		
	\end{minipage}%
	\begin{minipage}{0.50\linewidth}
		
		\centering{
			
			\includegraphics[keepaspectratio, width=\textwidth]{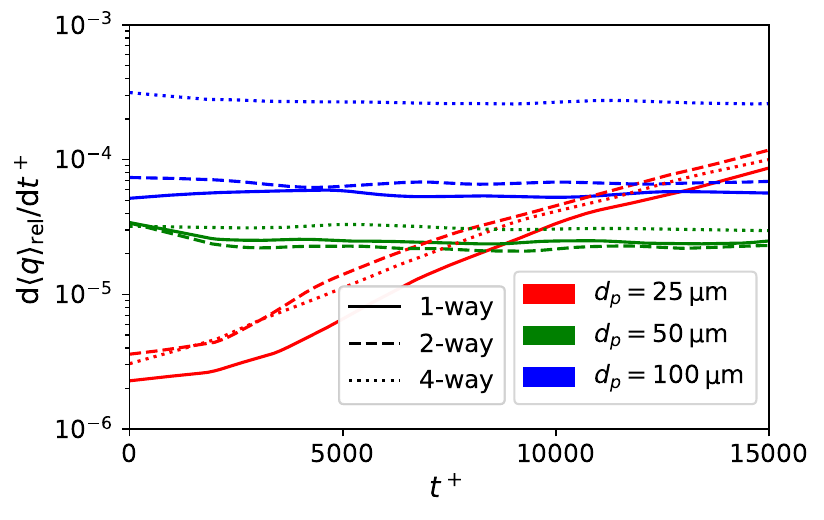}
			
		}
		
		\subcaption{\label{fig-charging-coupling-2}}
		
	\end{minipage}%
	
	\caption{\label{fig-charging-coupling} (a) Average relative particle charge \(\langle q\rangle_\text{rel}\) and (b) charging rate \(\langle q \rangle_\text{rel}/\mathrm{d}t^+\) over time for different particle sizes and particle-fluid coupling complexities for DNS of \(\Rey_\tau=180\).}
	
\end{figure}

Figure~\ref{fig-charging-coupling-1} shows the temporal evolution of the
relative mean charge \(\langle q \rangle_\text{rel}\) as a function of
\(t^+\), while figure~\ref{fig-charging-coupling-2} presents the
corresponding charging rate
\(\langle q \rangle_\text{rel}/\mathrm{d}t^+\). Overall, the charging behaviour is dominated by particle size rather than by the coupling mechanism. For particles of \(d_p = 25\) and
\(50\,\unit{\um}\), only minor differences arise between 1-way, 2-way,
and 4-way coupling. Specifically, \(25\,\unit{\um}\) particles exhibit a
slightly faster charge build-up when switching from 1-way to 2-way
coupling, whereas 4-way coupling causes almost no additional change. For
particles of \(d_p = 50\,\unit{\um}\), the opposite trend is observed:
the transition from 1-way to 2-way coupling barely affects the charging
dynamics, while 4-way coupling yields a modest increase in charging
rate. In contrast, \(100\,\unit{\um}\) particles experience a more
notable enhancement in charging rate when switching from 1-way to 2-way
coupling (by approximately 22 \%), followed by an even stronger increase
of about 296 \% when activating 4-way coupling, as illustrated in
figure~\ref{fig-charging-coupling-2}.

Figure~\ref{fig-charging-coupling} shows a longer time interval up to
\(t^+=15\,000\) than
figure~\ref{fig-charging-profile-oneway-condenser-ssm-e13-1}, although
the system is still in the early charging stage with the average charge
remaining below \(1\,\%\) of \(q_\text{sat}\). The focus of this work is placed on this early phase of charging, when the most significant differences between configurations occur, whereas the charging rates decay once saturation charge is approached \citep{Ozler2025}.
Within this investigated time, no significant temporal variation in the charging rate is observed
for particles of \(d_p = 50\,\unit{\um}\) and \(100\,\unit{\um}\). However,
\(d_p = 25\,\unit{\um}\) particles exhibit an almost logarithmic
increase in charging rate that gradually weakens as saturation becomes
imminent. As a consequence, these small particles eventually surpass the
charging rate of particles with \(d_p = 50\,\unit{\um}\) for
\(t^+ \gtrapprox 8000\) and even exceed the rate of \(100\,\unit{\um}\)
particles under 1-way and 2-way coupling once \(t^+ \gtrapprox 13\,000\).
This behaviour originates from electrostatic forces---without them, the
effect does not occur (not shown here).

\begin{figure}
	
	\begin{minipage}{0.50\linewidth}
		
		\centering{
			
			\includegraphics[keepaspectratio, width=\textwidth]{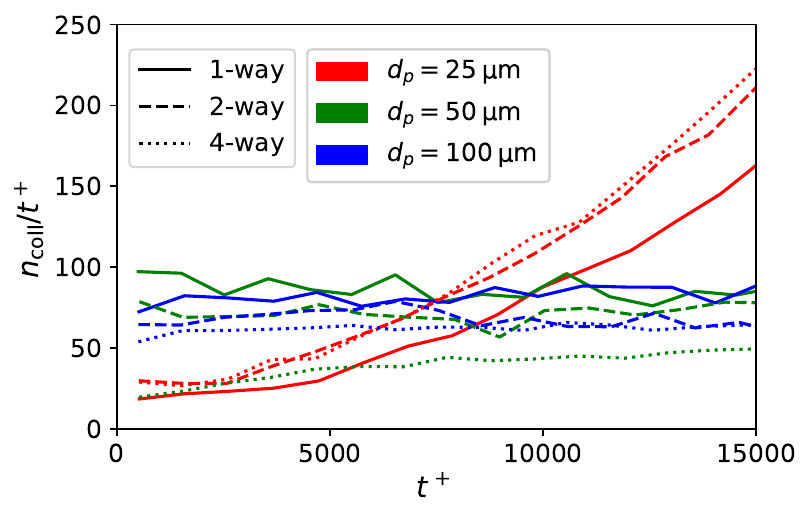}
			
		}
		
		\subcaption{\label{fig-charging-coupling-collisions-1}}
		
	\end{minipage}%
	\begin{minipage}{0.50\linewidth}
		
		\centering{
			
			\includegraphics[keepaspectratio, width=\textwidth]{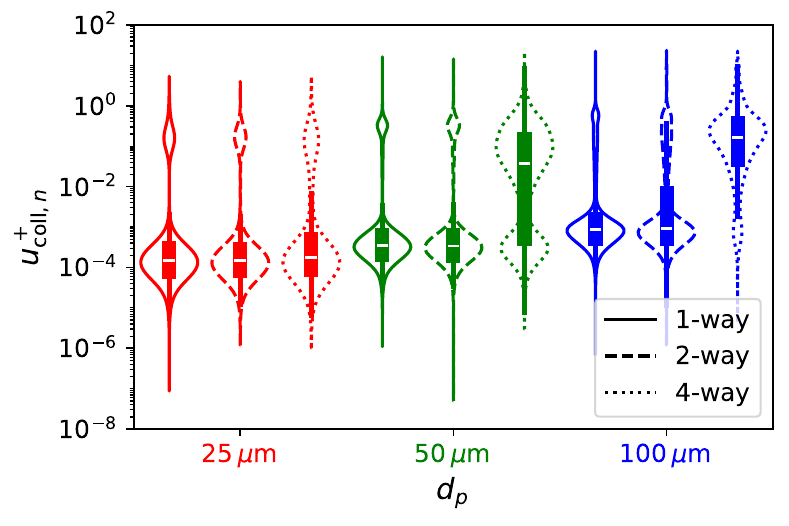}
			
		}
		
		\subcaption{\label{fig-charging-coupling-collisions-2}}
		
	\end{minipage}%
	
	\caption{\label{fig-charging-coupling-collisions}
    (a) Particle-wall collision rates \(n_\text{coll} / t^+\) and (b) distributions	of the wall-normal collision velocity \(u_{\text{coll}, n}^+\) over time for different particle sizes and coupling complexities for DNS of \(\Rey_\tau=180\).}
	
\end{figure}

To gain further insight into the charging rates and their underlying
physical mechanisms, the two principal drivers of charge transfer are
shown in figure~\ref{fig-charging-coupling-collisions}: (a) the temporal
evolution of the collision rate and (b) the collision velocity
distribution of uncharged particles. The time history of the collision
rate exhibits a qualitative resemblance to the charging rate, indicating
that the observed behaviour is primarily governed by the collision
frequency. Particles of \(d_p = 50\,\unit{\um}\) and \(100\,\unit{\um}\) experience
an almost constant collision rate throughout the analysed period,
whereas particles of \(d_p = 25\,\unit{\um}\) show an increase by
roughly one order of magnitude within the same interval.

For particles with \(St > 1\), the collision rate decreases from 1-way to 2-way and onwards to 4-way coupling, which
is consistent with the reduction of particle concentration in the very
near-wall region shown in
figure~\ref{fig-dns-sizes-coupling-particledensity}. In contrast,
particles of \(d_p = 25\,\unit{\um}\) experience an increase in
collision frequency when switching from 1-way to 2-way coupling, again
in agreement with the corresponding concentration distribution.
Interestingly, although 4-way coupling reduces the near-wall
concentration for these small particles, the collision rate remains
nearly unchanged relative to the 2-way case. This suggests that
particle--particle reflections in the near-wall region compensate for
the reduced particle density by redirecting off-bouncing particles back
toward the wall, thereby maintaining the collision frequency.

The rise in collision rate for the \(d_p = 25\,\unit{\um}\)
particles is an electrostatic effect. Larger particles do not show this behaviour, as their electric Stokes number is too small for electrostatic forces to exert a notable influence on their dynamics. The increased collision frequency is therefore attributable exclusively to charged particles, which exhibit a heightened probability
of rebounding collisions.
Two electrostatic mechanisms are possible: repulsion from already charged particles in the channel core,
and attraction toward the wall, which---as a grounded
conductor---maintains zero electric potential. The latter mechanism is dominant, as simulations employing only Coulomb forces between particles do not reproduce the increased collision rate; only the Gauss-law component of
the hybrid charging model produces such behaviour. It describes the effect of an image charge generated in the conductive wall which attracts the particles to the wall \citep{Ozler2025}.

Since the time evolution of particle charging rate and particle--wall
collision frequency match almost perfectly for \(d_p = 25\,\unit{\um}\), the coupling mechanism has no significant influence on the collision, as confirmed by figure~\ref{fig-charging-coupling-collisions-2}.
The collision-velocity distribution, shown for uncharged particles averaged over the charging period, exhibits no
significant temporal variation. Overall, the distribution is bimodal: the
lower mode corresponds to particles grazing the wall with small
tangential contact velocities, whereas the higher mode arises from
particles impinging during sweeps originating from the channel centre.

For \(d_p = 25\,\unit{\um}\), the collision-velocity distributions for
1-way and 2-way coupling are nearly identical, as the particle inertia
is too small to transport substantial momentum from the bulk toward the
wall. When 4-way coupling is enabled, the number of high-speed impacts and the distribution's overall spread slightly increases.
This increase is due to particle--particle collisions in the near-wall region that redirects previously rebounding particles back toward the wall. For \(d_p = 50\,\unit{\um}\), the transition from 1-way to 2-way
coupling again produces only minor changes. However, under 4-way
coupling the collision rate decreases while the charging rate increases
markedly. This increase is  linked to a rise of the average collision velocity by about two orders of magnitude. The
corresponding distribution demonstrates a shift in the dominant collision mechanism, with high-speed impacts overtaking the
low-speed grazing events.

For the largest particles, \(d_p = 100\,\unit{\um}\), the same
qualitative trends are observed, but now even 2-way coupling produces
noticeable changes, as the particle inertia becomes sufficient to
transport significant momentum toward the channel wall. Under 4-way
coupling, the low-speed grazing events nearly vanish, consistent with
the reduction in particle concentration at \(y^+ \approx d_p^+/2\)
in figure~\ref{fig-dns-sizes-coupling-particledensity}. The increased diameter raises the likelihood that near-wall particles are displaced by particles arriving from the channel core, and this effect becomes so dominant that the reduced collision frequency is fully outweighed by the substantially higher charge transfer per collision, leading to an overall increase in the charging rate of the particle ensemble.

\subsection{Extension toward LES}\label{validation-of-les-model}

\begin{figure}
	
	\begin{minipage}{0.50\linewidth}
		
		\centering{
			
		\includegraphics[keepaspectratio, width=\textwidth]{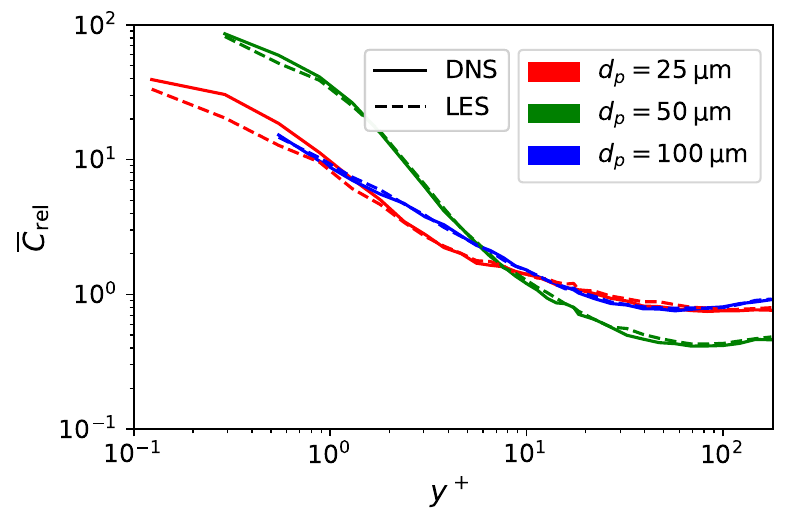}
			
		}
		
		\subcaption{\label{fig-dns-vs-les-sizes-4way-particledensity}Relative particle density \(\overline{C}_\text{rel}\).}
		
	\end{minipage}%
	\begin{minipage}{0.50\linewidth}
		
		\centering{
			
		\includegraphics[keepaspectratio, width=\textwidth]{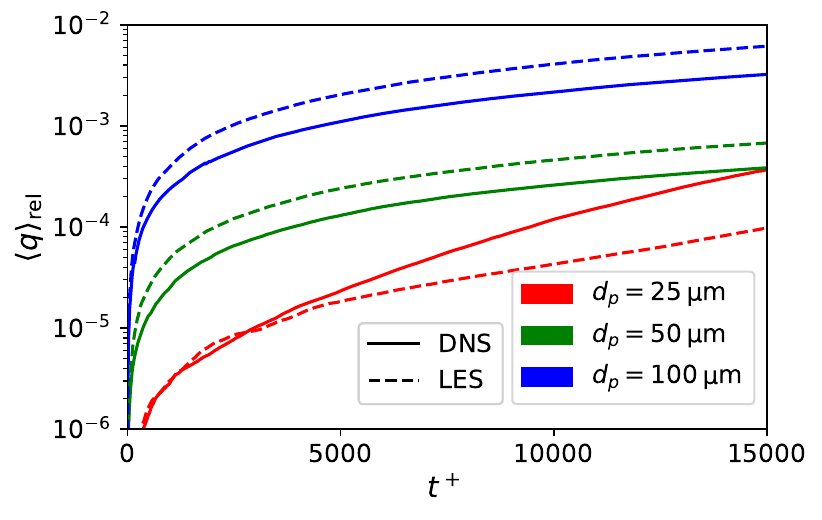}
			
		}
		
		\subcaption{\label{fig-dns-vs-les-sizes-4way-charging-e13}Averaged relative
			particle charge \(\langle q \rangle_\text{rel}\).}
		
	\end{minipage}%
	
	\caption{\label{fig-dns-vs-les-sizes-4way-charging-e13-total}(a) Relative
		particle density based on the particle center \(\overline{C}_\text{rel}\) over wall-normal
		coordinate \(y^+\) for uncharged particles of different sizes.
        (b) Averaged relative particle charge \(\langle q \rangle_\text{rel}\) over time for
		different particle sizes and for DNS and LES of \(\Rey_\tau=180\).}
	
\end{figure}

This section validates the LES model at a friction Reynolds number of
\(\Rey_\tau = 180\) as a preparation for the subsequent analysis at higher
Reynolds numbers. In a precursor study, we determined an appropriate grid resolution, see Section~\ref{sec-apx-grid-study-les}.
In addition, different subgrid-scale closure models were evaluated, see Section~\ref{sec-apx-les-models}, where the WALE closure model performed best. In general, provided that the grid resolution is
sufficient, LES reproduces the DNS results with good accuracy.
Figure~\ref{fig-dns-vs-les-sizes-4way-particledensity} compares the
wall-normal particle concentration profiles obtained from LES and DNS.
The agreement is best for the largest particles and decreases with
decreasing Stokes number. In particular, LES underpredicts the near-wall
particle concentration, with maximum deviations of approximately
\(10\,\%\) for \(d_p = 25\, \unit{\um}\), \(5\,\%\) for
\(d_p = 50\, \unit{\um}\), and \(1\,\%\) for
\(d_p = 100\, \unit{\um}\). This trend is due to LES not resolving the smallest turbulent scales, and
since small particles are most strongly influenced by these flow structures, the reduced representation of near-wall turbulence diminishes turbophoretic transport, leading to lower predicted concentrations at the wall \citep{Kuerten2004}.

Figure~\ref{fig-dns-vs-les-sizes-4way-charging-e13} presents the
comparison between DNS and LES with respect to the temporal evolution of
the relative particle charge. Overall, the LES reproduces the DNS. During the initial charging phase, the LES predictions for particles of \(d_p = 25\,\unit{\um}\) agree well with the DNS. With
increasing particle diameter, the LES begins to overpredict the charging, and this deviation grows systematically with particle size. For \(t^+ \gtrapprox 4000\), the LES results for \(d_p = 50\,\unit{\um}\)
and \(d_p = 100\,\unit{\um}\) continue to exceed the DNS values because
the LES charging rate remains approximately constant. The same behaviour
is observed for the \(d_p = 25\,\unit{\um}\) particles; however, since
the DNS exhibits an increase in charging rate in this regime, the LES
increasingly underpredicts the DNS for this smallest particle class.

\begin{figure}
	
	\begin{minipage}{0.50\linewidth}
		
		\centering{
			
			\includegraphics[keepaspectratio, width=\textwidth]{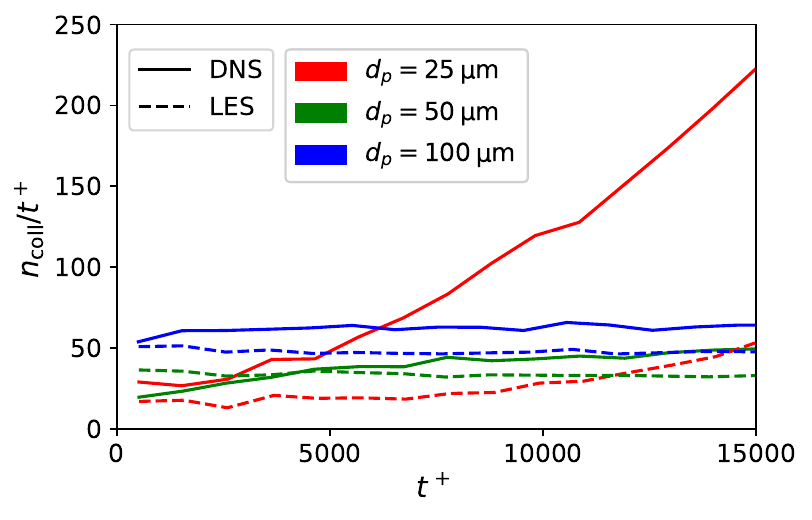}
			
		}
		
		\subcaption{\label{fig-dns-vs-les-sizes-4way-charging-collision-1}Particle
			collision rate \(n_\text{coll} / t^+\) over time \(t^+\).}
		
	\end{minipage}%
	\begin{minipage}{0.50\linewidth}
		
		\centering{
			
			\includegraphics[keepaspectratio, width=\textwidth]{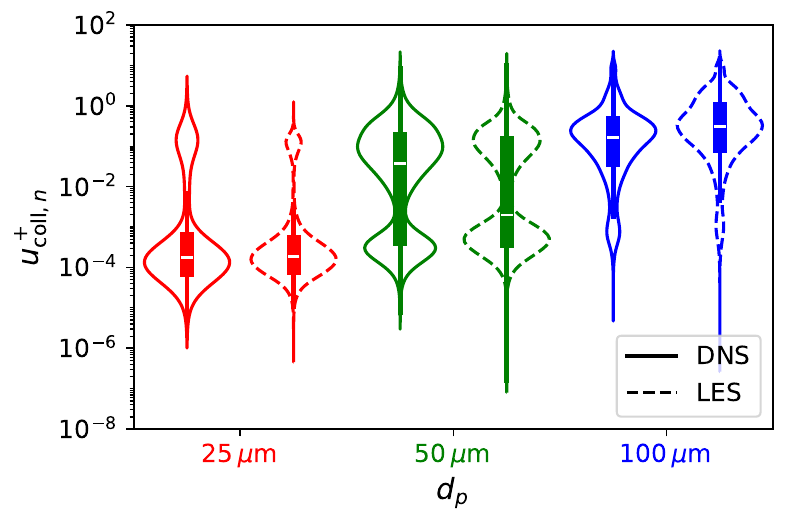}
			
		}
		
		\subcaption{\label{fig-dns-vs-les-sizes-4way-charging-collision-2}Particle wall-normal colliosion velocity
			\(u_{\text{coll}, n}^+\).}
		
	\end{minipage}%
	
	\caption{\label{fig-dns-vs-les-sizes-4way-charging-collision}Comparison
		of (a) particle-wall collision rate \(n_\text{coll} / t^+\) and (b)
		distributions of the collision velocity \(u_{\text{coll}, n}^+\) for
		different particle sizes and for DNS and LES of \(\Rey_\tau=180\).}
	
\end{figure}

Further insights into these discrepancies are provided by the comparison
of particle--wall collision rate and collision velocity shown in
figure~\ref{fig-dns-vs-les-sizes-4way-charging-collision}. For the
smallest particles of \(d_p = 25\,\unit{\um}\), the dominant source of
deviation lies in the collision rate. The electrostatic mechanism
associated with the grounded wall, arising from the Gauss-law component
of the electrostatic model, is significantly weaker in the LES. This
behaviour is directly linked to grid resolution: Gauss's law is Eulerian
and therefore grid dependent; the coarser LES grid resolves gradients in the electric potential less sharp; weaker potential gradients produce weaker electrostatic forces; and, as a result, particles are more likely to rebound from the wall, re-enter the turbulent flow, and be convectively transported away from the wall.
Consequently, the LES underestimates the electrostatically enhanced collision frequency.

For the intermediate particle size, \(d_p = 50\,\unit{\um}\), the DNS
shows a moderate increase in collision rate over time, and this effect
is partly reproduced by the LES. However, an additional mechanism
becomes relevant: the LES predicts a higher frequency of high-velocity
impacts, which increases the average charge transfer per collision and
contributes to the overall overprediction of the charging rate.

For the largest particles of \(d_p = 100\,\unit{\um}\), the LES even
underpredicts the collision rate relative to the DNS. The resulting
overestimation of the charging rate is therefore entirely attributable
to an increase in the collision velocity. Both the number of high-speed impacts and the probability of very high impact velocities increase in the LES. These energetic collisions are associated with particles
arriving from the channel centre. In the LES, such particles slow less down before hitting the wall. Several
mechanisms contribute to this discrepancy. First, LES inherently
filters the fluid velocity field, removing small-scale near-wall
turbulent structures that in DNS scatter particles and reduce their
velocity prior to impact. It has been shown that even well-resolved LES
may not accurately reproduce near-wall particle accumulation or
preferential segregation due to the absence of these fine scales,
leading to errors in particle velocity and concentration statistics near
the wall \citep{Marchioli2008}. Second, the filtering introduced by LES
alters the forces acting on particles, and the associated filtering
error is largest in the buffer region where near-wall interactions are
most critical; this can result in less attenuation of particle motion
toward the wall \citep{Bianco2012}. Finally, differences in resolved
turbulence spectra between LES and DNS influence turbophoresis and local
particle concentration gradients, further affecting collision statistics
\citep{Kuerten2016}. Each of these effects merits further investigation
to quantify their relative influence on particle--wall impact velocities
and associated charging.

In summary, LES provides an adequate approach for modelling particle
charging with substantially reduced computational cost, while capturing
the correct qualitative trends. However, LES may overpredict the charging rate for larger particles, which is
conservative from a safety perspective. For small particles, the
rebounding effect of charged particles is underestimated. Overall, LES model is validated, allowing
for the subsequent investigation of higher carrier-flow Reynolds numbers in the next section.

\subsection{Effect of Reynolds number}\label{sec-re-study}

Using the validated LES, we increased the Reynolds number in the 4-way coupled LES to \(\Rey_\tau = 300, 395,\) and \(550\). Owing to the higher centreline velocity at increasing \(\Rey_\tau\), the Stokes number \(St\)
also increases (see table~\ref{tbl-particle-nondimensional}). In
addition, the particle diameter in wall units becomes larger. Both
effects imply that the particle dynamics are governed by
the larger turbulent scales.

\begin{figure}
	
	\begin{minipage}{\linewidth}
		
		\centering{
			
			\includegraphics[keepaspectratio, width=\textwidth]{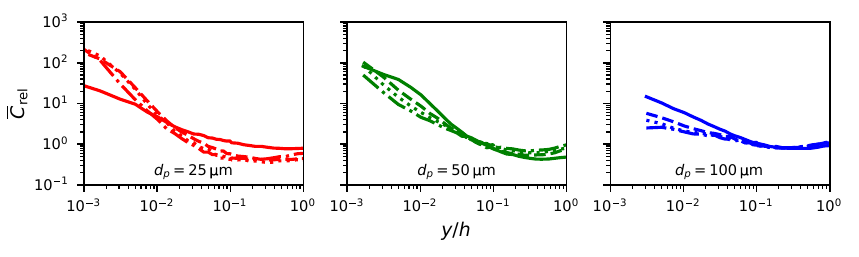}
			
		}
		
		\subcaption{\label{fig-180vs300-4way-particle-density-1}}
		
	\end{minipage}%
	\newline
	\begin{minipage}{\linewidth}
		
		\centering{
			
			\includegraphics[keepaspectratio, width=\textwidth]{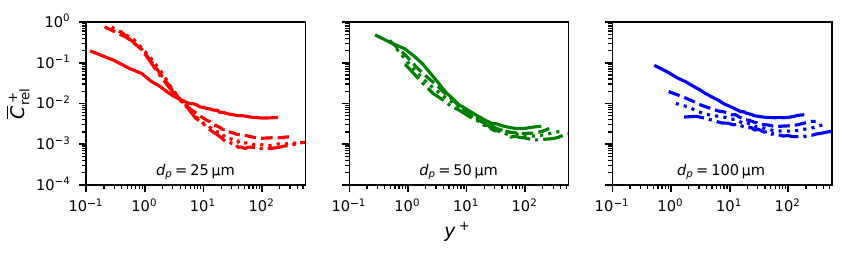}
			
		}
		
		\subcaption{\label{fig-180vs300-4way-particle-density-2}}
		
	\end{minipage}%
	\newline
	
	\begin{minipage}{\linewidth}
		
		\centering{
			
			\includegraphics[keepaspectratio, width=\textwidth]{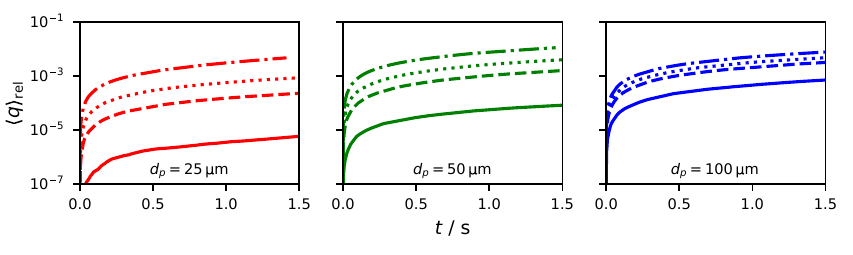}
			
		}
		
		\subcaption{\label{fig-180vs300-4way-charging-e13-1}}
		
	\end{minipage}%
	\newline
	\begin{minipage}{\linewidth}
		
		\centering{
			
			\includegraphics[keepaspectratio, width=\textwidth]{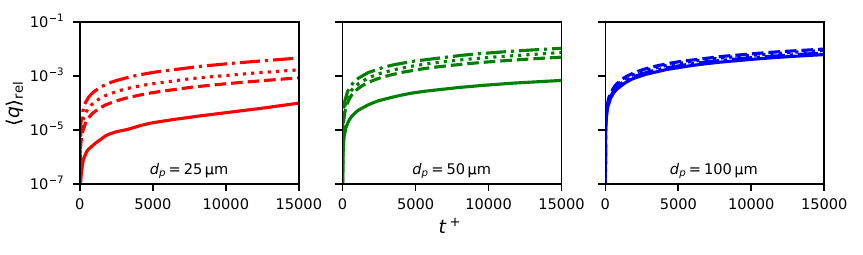}
			\includegraphics[keepaspectratio, width=.7\textwidth]{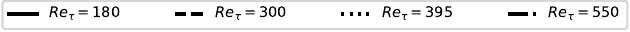}
		}
		
		\subcaption{\label{fig-180vs300-4way-charging-e13-2}}
		
	\end{minipage}%
	
	\caption{\label{fig-180vs300-4way-charging-e13}
    LES for particles of different sizes and friction Reynolds numbers.
    Relative density of uncharged particles \(C_\text{rel}\) plotted over wall-normal coordinate (a) in outer units and (b) inner/wall
		units).
    Averaged relative particle charge \(\langle q \rangle_\text{rel}\) plotted over time in	(c) seconds \(t\) and (d) wall units \(t^+\).}
	
\end{figure}

Figure~\ref{fig-180vs300-4way-particle-density-1} presents the
wall-normal particle concentration profiles for the four investigated
Reynolds numbers and the three particle sizes, scaled with the outer
length scale (channel height \(h\)). Overall, the particle distribution
is more influenced by the particle diameter than by the
Reynolds number. For the smallest particles (\(d_p = 25\,\unit{\um}\)),
an increase in \(\Rey_\tau\) results in a higher particle concentration
near the wall. This trend appears asymptotic, with the most pronounced
rise occurring between \(\Rey_\tau = 180\) and \(\Rey_\tau = 300\),
corresponding to an increase of approximately \(570\,\%\). In contrast,
for the larger particles (\(d_p = 50\,\unit{\um}\) and
\(d_p = 100\,\unit{\um}\)), the opposite behaviour is observed: the
near-wall particle concentration decreases with increasing Reynolds
number. These observations are consistent with the findings of
\citet{Bernardini2014}, who reported that the particle concentration in
the outer layer scales with outer units, particularly for \(St = 1\) and
\(St \ge 500\). In the range of \(10 \ge St \ge 100\), no scaling with
the outer layer is reported, which is both in agreement with the trends
shown in figure~\ref{fig-180vs300-4way-particle-density-1}.

For the inner layer, \citet{Sardina2013} and \citet{Bernardini2014}
proposed a scaling with inner units, especially for intermediate Stokes
numbers (\(10 \lessapprox St \lessapprox 25\)). Following this approach,
\citet{Bernardini2014} introduced a particle concentration in wall
units, \(C^+_\text{rel} = C_\text{rel} / \Rey_\tau\), which is shown in
figure~\ref{fig-180vs300-4way-particle-density-2}. The figure reveals
similar qualitative trends to those in
figure~\ref{fig-180vs300-4way-particle-density-1}: for small particles
(\(d_p = 25\,\unit{\um}\)), increasing \(\Rey_\tau\) enhances the wall
concentration, whereas for larger particles \(d_p \ge 50\,\unit{\um}\))
the concentration decreases with increasing Reynolds number. The latter
trend is consistent with the observations of \citet{Bernardini2014},
while the increase in concentration for \(d_p = 25\,\unit{\um}\) was not
reported in their study, as they considered only \(St \ge 1\). Moreover,
the expected collapse for \(10 \lessapprox St \lessapprox 25\) is not
observed here. This deviation may be attributed to 4-way coupling and higher particle concentrations,
in contrast to the 1-way coupled configuration of
\citet{Bernardini2014}. The decrease in near-wall concentration with
increasing Reynolds number can be due to a smaller turbophoretic
drift as the fraction of inner layer to channel height decreases.

Figure~\ref{fig-180vs300-4way-charging-e13-1} shows the charging
behaviour for different Reynolds numbers, where the temporal evolution
of the charge is plotted in seconds. For all particle sizes, an increase
in Reynolds number leads to a higher charging rate, which is in
accordance with experimental \citep{Schwindt2017, Nifuku2003} and
numerical observations \citep{Grosshans2016, Watano2003, Ceresiat2019}.
This effect is most pronounced for the smallest particles
(\(d_p = 25\,\unit{\um}\)), where the charging rate increases by
approximately two orders of magnitude as \(\Rey_\tau\) rises from \(180\)
to \(550\). For the largest particles (\(d_p = 100\,\unit{\um}\)), the
effect is weaker, with only about one order of magnitude
increase in charging rate. As a result, the highest charging rate is
observed for medium-sized particles (\(d_p = 50\,\unit{\um}\)) at
\(\Rey_\tau = 550\). When the charge is plotted over wall units, \(t^+\),
as shown in figure~\ref{fig-180vs300-4way-charging-e13-2}, an asymptotic
behaviour with increasing \(\Rey_\tau\) becomes apparent, particularly for
the larger particles. This tendency may facilitate the prediction of the
charging rate at higher Reynolds numbers. However, the scaling appears
to depend on particle size, as no collapse is observed for the smaller
particles. Verification at higher Reynolds numbers is therefore required
as this concept has not been previously reported in the literature.

\begin{figure}
	
	\begin{minipage}{0.50\linewidth}
		
		\centering{
			
			\includegraphics[keepaspectratio, width=\textwidth]{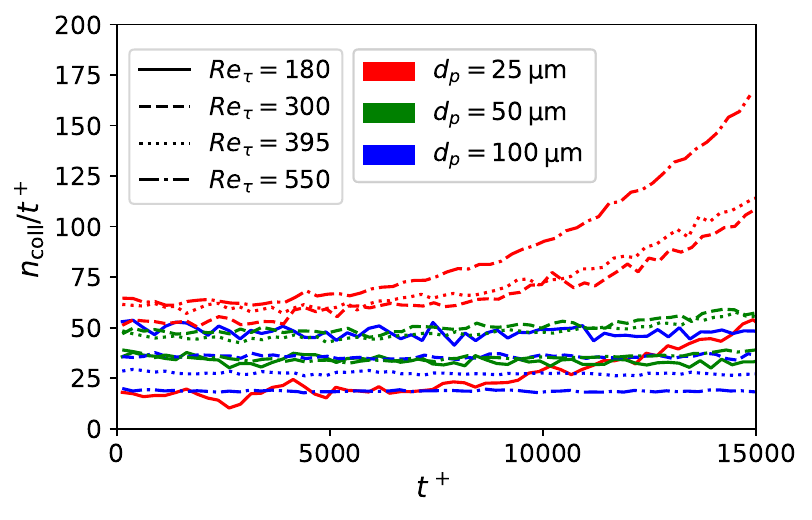}
			
		}
		
		\subcaption{\label{fig-180vs300-4way-collision-e13-1}}
		
	\end{minipage}%
	\begin{minipage}{0.50\linewidth}
		
		\centering{
			
			\includegraphics[keepaspectratio, width=\textwidth]{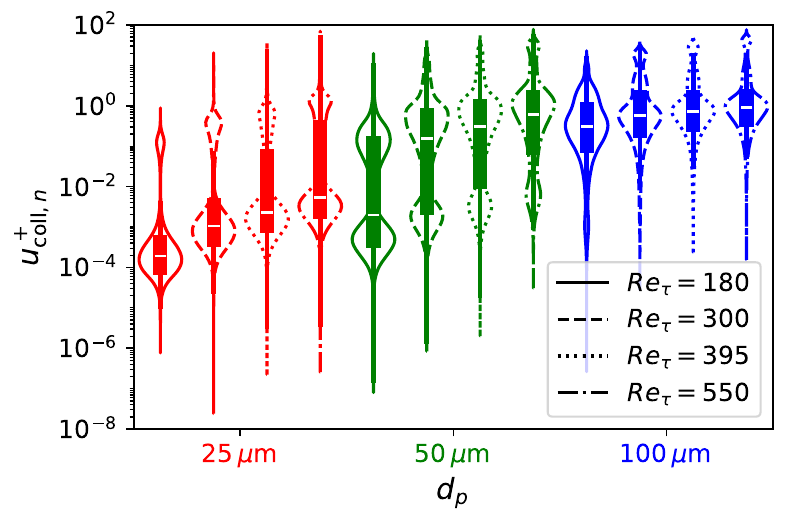}
			
		}
		
		\subcaption{\label{fig-180vs300-4way-collision-e13-2}}
		
	\end{minipage}%
	
	\caption{\label{fig-180vs300-4way-collision-e13}
    Four-way coupled LES for particles of different sizes and friction Reynolds numbers.
    (a) Particle collision rate \(n_\text{coll} / t^+\).
    (b) Distribution of the wall-normal collision velocity \(u_{\text{coll}, n}^+\).}
	
\end{figure}%

To further assess the influence of Reynolds number on particle charging,
figure~\ref{fig-180vs300-4way-collision-e13} shows the particle--wall
collision rate and the corresponding collision velocity. In agreement
with the trends in figure~\ref{fig-dns-vs-les-sizes-4way-charging-collision}, smaller
particles exhibit an increase in collision frequency and charging rate
over time (the latter not shown). These increases become stronger for higher Reynolds numbers. As discussed above, the enhanced collision rate is associated with recurring particle--wall
interactions driven by the image charge force. A possible explanation
for the Reynolds-number dependence is the higher collision velocity at
larger Reynolds numbers, which accelerates particle charging and thus
strengthens the image charge and the resulting electrostatic attraction
towards the wall. In addition, the increased grid resolution relative to
the particle diameter resolves steeper gradients of the electric
potential. The relative contribution of these effects
cannot be isolated on the basis of the present results. For the largest
particles (\(d_p = 100\,\unit{\um}\)), the collision rate per unit wall
time decreases with increasing Reynolds number. This trend is consistent
with the reduced near-wall particle concentration observed at higher
Reynolds numbers, an effect that is particularly pronounced for large
particles.

The mean collision velocity increases with Reynolds number for all
particle sizes, with the strongest increase for the smallest
particles. The increased momentum of large-scale turbulent eddies for higher Reynolds numbers transports particles
towards the wall. For particles with diameters of \(d_p = 25\,\unit{\um}\) and especially for \(50\,\unit{\um}\), a shift in the collision statistics from low-velocity to high-velocity impacts is observed with increasing Reynolds number. This
shift aligns with the increase in turbulence intensity at higher
Reynolds numbers. Stronger sweeps reduce particle residence times
in the near-wall region, thereby suppressing low-velocity grazing
collisions.

\begin{figure}
	
	\centering{
		
		\begin{minipage}{0.50\linewidth}
		
		\centering{
			
			\includegraphics[keepaspectratio, width=\textwidth]{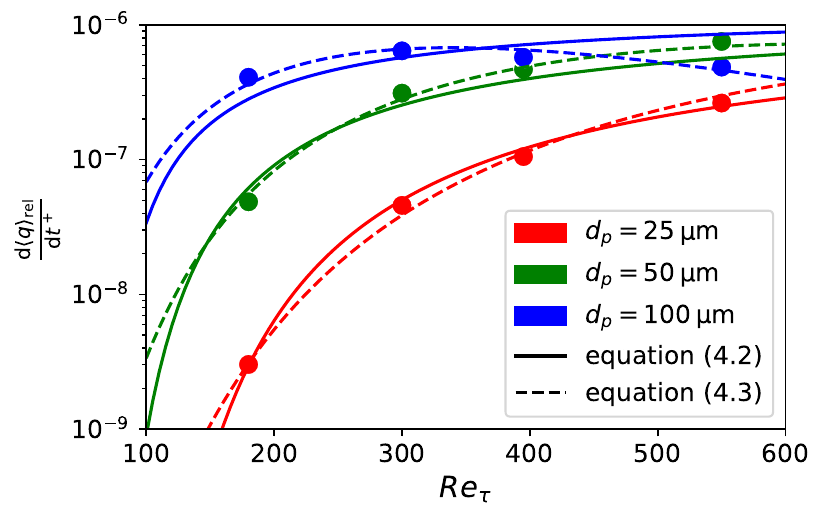}
			
		}
		
	\end{minipage}%
		
	}
		
	\caption{\label{fig-180vs300-4way-charging-rate-correlation}Correlation
		between friction Reynolds number and average charging rate
		\(d\langle q \rangle_\text{rel}/dt\) for different particle sizes.}
	
\end{figure}%

As a final step, figure~\ref{fig-180vs300-4way-charging-rate-correlation} shows the
charging rate as a function of \(\Rey_\tau\) for the different particle
sizes. In general, the charging rate exhibits a temporal evolution, most
notably for small particles, for which a gradual increase is observed
over time (see figure~\ref{fig-charging-coupling-2}). To eliminate this
transient effect, the analysis is restricted to \(t^+ < 4000\), where
the charging rate is approximately constant for all cases. Consequently,
the extracted charging rates exclude the influence of enhanced collision
frequencies caused by image-charge-induced recurring collisions, as well
as particle saturation effects that become relevant at later times.

According to \citet{Ozler2025}, the charging
rate reduces once the mean particle charge reaches approximately
\(20\)--\(40\,\%\) of the saturation charge, \(q_\text{sat}\).
The influence of increased collision rates is confined to charge levels
below approximately \(10\,\%,q_\text{sat}\), since particles undergoing
recurring collisions have already approached saturation. As a result,
the charging rates reported here are expected to overpredict the charge
accumulation when extrapolated towards \(q_\text{sat}\). All charging
rates are determined on a temporal basis expressed in wall units, as the
asymptotic behaviour is observed most clearly in this scaling. Overall,
an increase in both \(\Rey_\tau\) and particle diameter \(d_p\) leads to a
higher charging rate per unit of \(t^+\). However, the tendency towards
an asymptotic, Reynolds-number-independent charging rate becomes more
pronounced with increasing particle size. For the largest particles
considered (\(d_p = 100\,\unit{\um}\)), the charging rate even decreases
for \(\Rey_\tau > 300\). 

To quantify these trends and to enable predictive
estimates for other configurations, we propose two empirical relations obtained using symbolic regression \citep{Cranmer2023}, namely
\begin{equation}\phantomsection\label{eq-pysr-result-simple}{ \log_{10} \left( \frac{\text{d}\langle q  \rangle_\text{rel}}{\text{d}t^+} \right) = -5.89 - \frac{9070}{d_{p, \text{rel}} \left(\Rey_\tau - 42.5\right)} }\end{equation}
and
\begin{equation}\phantomsection\label{eq-pysr-result}{ \log_{10} \left( \frac{\text{d}\langle q  \rangle_\text{rel}}{\text{d}t^+} \right) = -2.18 \cdot 10^{-5} \, \Rey_\tau  d_{p, \text{rel}} - 4.64 - \frac{30360}{\Rey_\tau \left(d_{p, \text{rel}} + 4.92\right) + 2668} }\end{equation}
with \(d_{p, \text{rel}} = d_p /\left(50 \delta\right) \). 
Equation~(\ref{eq-pysr-result-simple}) provides a compact representation
of the charging-rate dependence at the expense of a slightly increased
fitting error, whereas equation~(\ref{eq-pysr-result}) yields a more
accurate description, but at the cost of increased complexity. The difference between the two formulations lies in their
behaviour at large Reynolds numbers. While
Equation~(\ref{eq-pysr-result-simple}) predicts an asymptotic approach
towards a maximum charging rate of approximately
\(\frac{{\text{d}\langle q \rangle_\text{rel}}}{{\text{d}t^+}} \approx 10^{-6} \text{C}\),
equation~(\ref{eq-pysr-result}) captures the reduction in charging rate
observed for large particles at high Reynolds numbers. An careful extrapolation
of equation~(\ref{eq-pysr-result}) towards higher Reynolds numbers suggests a similar decrease for
smaller particles as well. Based on this formulation, the maximum
charging rate is predicted to occur for particles with
\(d_p = 25\,\unit{\um}\) at \(\Rey_\tau \approx 1280\), for
\(d_p = 50\,\unit{\um}\) at \(\Rey_\tau \approx 665\), and for
\(d_p = 100\,\unit{\um}\) at \(\Rey_\tau \approx 340\). Beyond these
Reynolds numbers, the charging rate decreases again. Since these Reynolds numbers are relevant to a wide range of industrial applications, the underlying mechanisms responsible for this non-monotonic behaviour warrant further investigation.

The present investigation was initially motivated by contradictory
trends reported in the literature regarding the Reynolds-number
dependence of particle charging. The results presented here are
consistent with recent numerical findings by \citet{Grosshans2016} and
with the experimental observations of \citet{Schwindt2017}, both of
which report an increase in particle charge with increasing Reynolds
number. Furthermore, the results support the conclusions of
\citet{Jantac2024}, who demonstrated that particle trajectories are
strongly influenced by turbulent flow structures that intensify with
increasing Reynolds number, leading to higher particle--wall impact
velocities and, consequently, enhanced charging rates. In contrast, the
simulations of \citet{Tanoue2001} and \citet{Watano2006} do not resolve
small-scale turbulent fluctuations, which provides a plausible
explanation for the differing trends reported in those studies.

\section{Conclusion}\label{sec-Conclusion}

We presented the new CFD-DEM code triboFoam to simulate triboelectric charging in particle-laden
flows. The code is based on the open-source library OpenFOAM, enabling parallel simulations of complex geometries combined with many physical models and numerical schemes.
In the first step, triboFoam has been validated by DNS of a
turbulent channel flow versus the open-source particle tracking library
pafiX. The results show an excellent agreement between both codes
for the fluid statistics as well as the particle distribution.
Concerning the charging, a match could be obtained by selecting
appropriate grid resolutions and time step sizes. In this study, we
could see that triboFoam is stable in terms of grid and time
step size variations. Moving forward, we extended triboFoam to LES
of turbulent channel flows at different friction Reynolds numbers
\(\Rey_\tau\) ranging from \(180\) to \(550\) and three different particle
sizes \(d_p = 25\,\unit{\um}\), \(50\,\unit{\um}\), and
\(100\,\unit{\um}\). The validation of the LES model highlighted how
sensitive the particle kinematics are to turbulent flow and that the
smallest eddies significantly contribute to a higher charging rate. The
results concerning different Reynolds numbers show that with increasing
Reynolds number the particle-wall collision velocity increases leading
to a higher charging rate. Furthermore, smaller particles are more
affected by the increase in Reynolds number as they follow the turbulent
structures better being transported towards the wall with higher
velocities. In addition, we observed for the small particles that with
increasing Reynolds number the particle-wall collision frequency
increases due to recurring collisions driven by the image charge force.
Finally, we derived two versions of an empirical correlation for the
average charging rate as a function of friction Reynolds number and
particle size. These show that the normalised charging rate tends
either asymptotically towards a maximum value or even decreases again
for high Reynolds numbers. This behaviour should be investigated in
future studies, as these Reynolds numbers are relevant for industrial
applications.

\bibliographystyle{jfm}
\bibliography{bibliography}


\begin{bmhead}[Funding]
This work received financial support from the European Research Council
(ERC) under the European Union's Horizon 2020 research and innovation
programme (Grant Agreement No.~947606, Pow-FEct). The authors gratefully
acknowledge the computing time made available to them on the
high-performance computer ``Lise'' at the NHR centre NHR@ZIB. This
centre is jointly supported by the Federal Ministry of Education and
Research and the state governments participating in the NHR
(www.nhr-verein.de).
\end{bmhead}

\begin{bmhead}[Acknowledgement]
Additional computational resources were provided by
the PTB cluster, which the authors also gratefully acknowledge. In addition, the authors like to thank Jiri Polansky from ESI-OpenCFD for his valuable feedback during the implementation of triboFoam in the OpenFOAM framework. 
\end{bmhead}

\begin{bmhead}[Declaration of Interests]
The authors report no conflict of interest.
\end{bmhead}

\begin{bmhead}[Data availability statement]
The data in the present manuscript will be made available upon request.
\end{bmhead}

\begin{bmhead}[Use of artificial intelligence tools]
The authors used ChatGPT v5.2 solely for the purpose of language editing and grammatical refinement to improve the readability of the manuscript. All scientific content, data analysis, and conclusions were developed exclusively by the human authors, who remain fully accountable for the accuracy, integrity, and originality of the work
\end{bmhead}

\begin{appen}

\section{DNS: Grid and time step study}\label{sec-apx-dns-grid-study}

This subsection evaluates the influence of grid resolution in the
wall-normal direction and the time step length.
Figure~\ref{fig-charging-profile-oneway-condenser-grid-timestep-study-1}
compares three grid resolutions, 144, 244, and 344 cells in the
wall-normal direction, in terms of particle charging over time. The
meshes are identical for pafiX and triboFoam.
triboFoam exhibits stable behaviour with respect to grid
variations for both particle sizes. In contrast, pafiX shows an
increased charging rate for small particles of \(d_p = 25\,\unit{\um}\)
at higher grid resolutions due to a higher particle-wall collision rate
and velocity. For larger particles of \(d_p = 100\,\unit{\um}\), no
significant dependence on grid resolution is observed, indicating that
smaller Stokes numbers are more sensitive to grid resolution.
Figure~\ref{fig-charging-profile-oneway-condenser-grid-timestep-study-2}
illustrates the influence of time step length on particle charging.
Again, triboFoam demonstrates stable behaviour with a slight
increase in charging rate. In contrast, pafiX exhibits a
noticeable increase in charging rate for all particle sizes with
decreasing time step size. An increase in charging rate is reasonable as
it resolves higher particle-wall collision frequencies.

\begin{figure}
	
	\begin{minipage}{0.50\linewidth}
		
		\centering{
			
			\includegraphics[keepaspectratio, width=\textwidth]{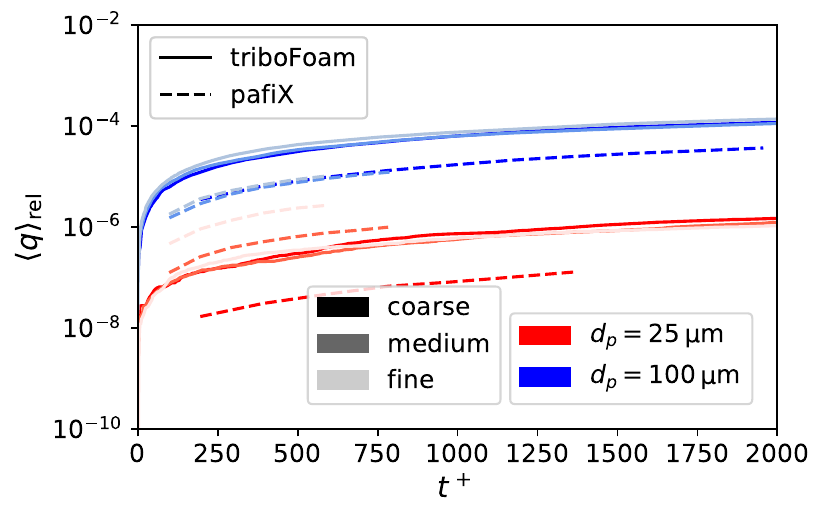}
			
		}
		
		\subcaption{\label{fig-charging-profile-oneway-condenser-grid-timestep-study-1}}
		
	\end{minipage}%
	\begin{minipage}{0.50\linewidth}
		
		\centering{
			
			\includegraphics[keepaspectratio, width=\textwidth]{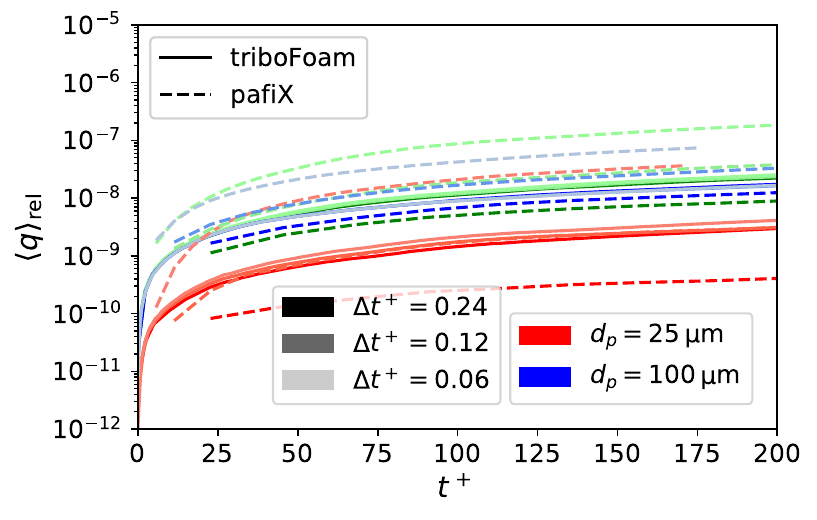}
		}
		
		\subcaption{\label{fig-charging-profile-oneway-condenser-grid-timestep-study-2}}
		
	\end{minipage}%
	
	\caption{\label{fig-charging-profile-oneway-condenser-grid-timestep-study}Averaged
		relative particle charge \(\overline{q}_\text{rel}\) over time for both
		solvers and two different particle sizes. Compared are in (a) different
		grid resolutions and in (b) different time step sizes. DNS of \(\Rey_\tau=180\), the particles
		are 1-way coupled to the fluid.}
	
\end{figure}%

To investigate this in more detail, and as the time step length has a
more significant effect on the charging behaviour than the grid
resolution,
figure~\ref{fig-charging-study-time-step-collision-velocity-frequency}
shows the mean particle-wall collision frequency and the distribution of
wall-normal particle collision velocity for different time step lengths.
It can be seen that with decreasing time step length, the mean
particle-wall collision frequency increases for both solvers. However,
triboFoam shows for the large particles almost no change and for
the medium sized particles the biggest change with about 0.7 orders of
magnitude. In contrast, pafiX shows a more consistent increase in
collision frequency with decreasing time step length for all particle
sizes in the range of 3 to 4 orders of magnitude. This suggests that
triboFoam is less sensitive to time step length variations for
larger particles compared to pafiX. In the very end, a realistic
collision frequency needs to be determined based on experiments to
choose an appropriate time step length for the simulations. The effect
of the time step length on the collision velocity is depicted in
figure~\ref{fig-charging-study-time-step-collision-velocity-frequency}.
For the particle size of \(d_p = 100 \, \unit{\um}\), no significant
trend can be observed for both solvers. However, for the small particle
size, triboFoam, again, shows almost no change with varying time
step lengths, while pafiX shows a trend towards lower collision
velocities with decreasing time step lengths. In addition, both solvers
differ in general in the distribution of the collision velocities, with
triboFoam showing a bimodal distribution with low and high speed
collisions and pafiX a more Gaussian distribution with the low
speed collisions. Similar to the collision frequency, this has to be
further investigated based on experimental data to choose an appropriate
time step length for the simulations.

\begin{figure}
	
	\begin{minipage}{0.50\linewidth}
		
		\centering{
			
			\includegraphics[keepaspectratio, width=\textwidth]{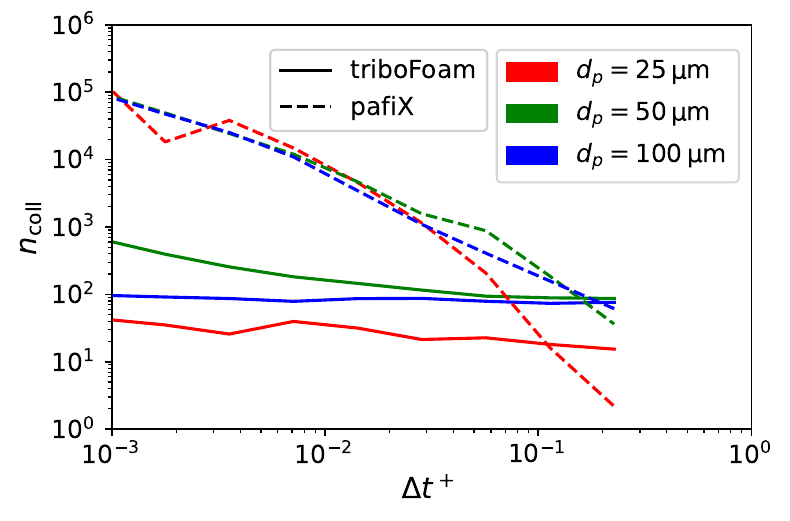}
			
		}
		
		\subcaption{\label{fig-charging-study-time-step-collision-velocity-frequency-1}}
		
	\end{minipage}%
	\begin{minipage}{0.50\linewidth}
		
		\centering{
			
			\includegraphics[keepaspectratio, width=\textwidth]{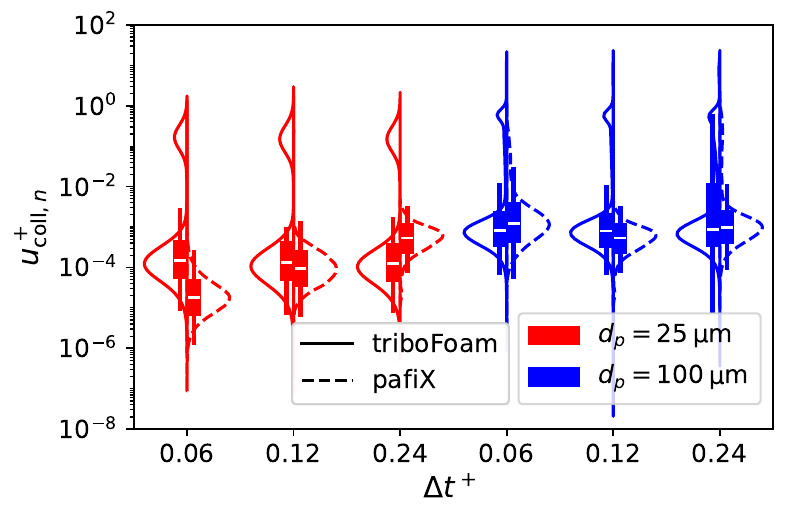}
			
		}
		
		\subcaption{\label{fig-charging-study-time-step-collision-velocity-frequency-2}}
		
	\end{minipage}%
	
	\caption{\label{fig-charging-study-time-step-collision-velocity-frequency}Effect
		of time step size on (a) the mean particle-wall collision frequency
		and (b) the distribution of wall-normal particle collision velocity
		\(u_{\text{coll}, n}^+\) for DNS of \(\Rey_\tau=180\).}
	
\end{figure}%

\section{LES: Comparison of closure models}\label{sec-apx-les-models}

Four different LES closure models were compared against a reference DNS:
Smagorinsky, \(k\)-equation (kEqn), dynamic \(k\)-equation (dykEqn), and
wall-adapting local eddy-viscosity (WALE). The comparison was carried
out at \(\Rey_\tau = 180\) using the smallest particle size, which is most
sensitive to variations in the fluid field. The grid corresponds to the
medium resolution described in Section~\ref{sec-apx-grid-study-les}.
Figure~\ref{fig-les-model-re180-4way-1} presents the wall-normal
velocity profiles for the different models. The kEqn model underpredicts
the bulk velocity, which appears to result from a delayed onset and
termination of the buffer layer. The other three models produce nearly
identical profiles, slightly overpredicting the bulk velocity. Agreement
with the DNS is very good up to \(y^+ < 10\) (within \(1\,\%\)), but the
velocity increases too rapidly beyond this region, indicating a delayed
onset of the logarithmic layer. All LES models underestimate the
velocity in the defect region (\(y^+ > 0.3\,\Rey_\tau\)), though this is
expected to have only a minor impact on the particle dynamics.
Figure~\ref{fig-les-model-re180-4way-2} compares the resulting particle
concentration profiles. The kEqn model predicts the near-wall
concentration most accurately, followed by the dykEqn and WALE models,
which show nearly identical trends, while the Smagorinsky model yields
the largest underprediction. Considering both the velocity and
concentration profiles, the dykEqn and WALE models perform best overall.
The WALE model was chosen for subsequent simulations, owing to its
\(16\,\%\) higher computational efficiency compared with the dykEqn
model.

\begin{figure}
	
	\begin{minipage}{0.50\linewidth}
		
		\centering{
			
			\includegraphics[keepaspectratio, width=\textwidth]{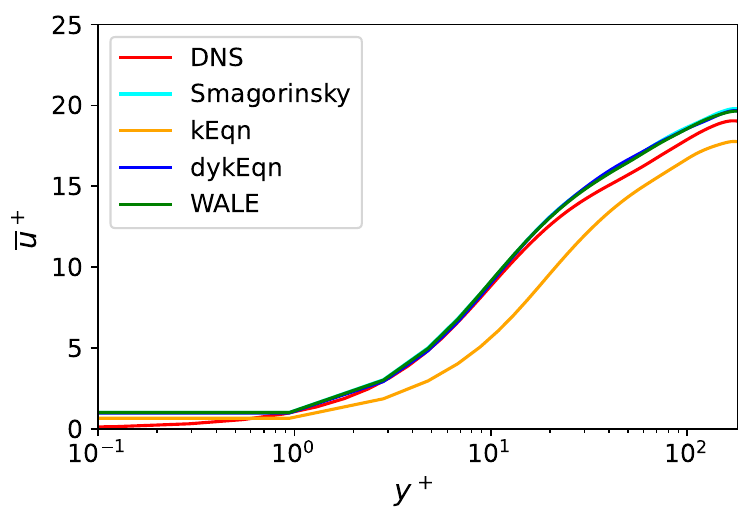}
			
		}
		
		\subcaption{\label{fig-les-model-re180-4way-1}}
		
	\end{minipage}%
	\begin{minipage}{0.50\linewidth}
		
		\centering{
			
			\includegraphics[keepaspectratio, width=\textwidth]{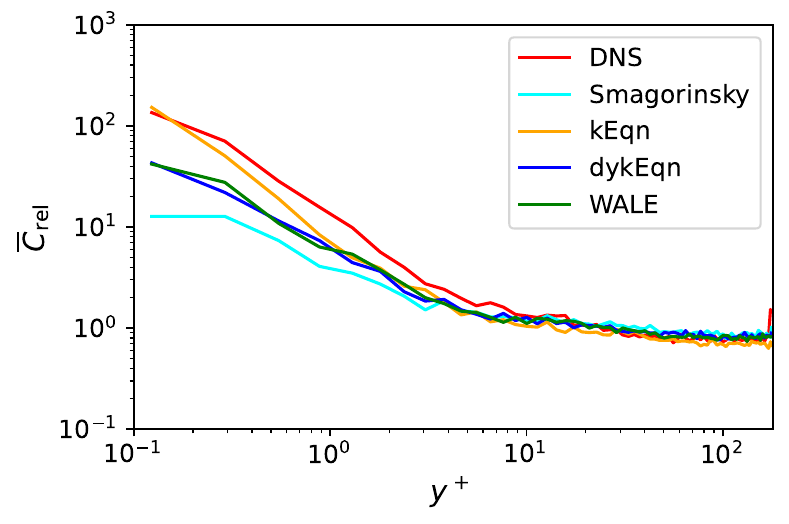}
			
		}
		
		\subcaption{\label{fig-les-model-re180-4way-2}}
		
	\end{minipage}%
	
	\caption{\label{fig-les-model-re180-4way}Effect of LES closure model on
		the velocity profile \(\overline{u}^+\) and particle concentration
		\(\overline{C}_\text{rel}\) plotted over wall-normal distance \(y^+\).
		All simulations for \(\Rey_\tau = 180\) with particles of a
		size with \(d_p = 25 \mu m\). The particles are 4-way coupled to the
		fluid.}
	
\end{figure}%

\section{LES: Grid study}\label{sec-apx-grid-study-les}

This section investigates the influence of grid resolution on the
velocity and particle concentration profiles in order to validate the
LES setup. The grid study compares three resolutions---referred to as
\emph{coarse}, \emph{medium}, and \emph{fine}---for friction Reynolds
numbers \(\Rey_\tau = 180\), \(300\), \(395\), and \(550\). The respective
wall-unit resolutions are \(\Delta x^+ = 39.5\), \(\Delta z^+ = 26.3\),
\(\Delta y^+_w = 1.9\), and \(\Delta y^+_w = 12.5\) (coarse),
\(\Delta x^+ = 19.75\), \(\Delta z^+ = 13.16\), \(\Delta y^+_w = 0.96\),
and \(\Delta y^+_w = 6.25\) (medium), and \(\Delta x^+ = 9.88\),
\(\Delta z^+ = 6.58\), \(\Delta y^+_w = 0.47\), and
\(\Delta y^+_w = 6.25\) (fine). At higher \(\Rey_\tau\), additional grid
variations were examined to reduce computational cost. The smallest
particle size, \(d_p = 25\,\unit{\um}\), was chosen as it is most
sensitive to the fluid field. DNS data at the corresponding Reynolds
numbers serve as reference: in-house DNS for \(\Rey_\tau = 180\) and
\(300\), and literature data for \(\Rey_\tau = 395\) and \(550\)
\citep{Abe2009, Lee2015}. For the latter two cases, only the velocity
profiles are compared since the DNS data are single-phase; however, the
influence of four-way coupling on the mean velocity is expected to be
minor due to the low particle concentration.

Figure~\ref{fig-dns-vs-les-re180-4way-gridstudy-1} shows that, at
\(\Rey_\tau = 180\), the LES velocity profile approaches the DNS reference
with increasing grid resolution. Coarser grids overpredict the bulk
velocity, while all grids accurately capture the viscous sublayer.
Beyond \(y^+ > 10\), the velocity rises too rapidly, delaying the onset
of the logarithmic layer; this discrepancy diminishes with increasing
resolution, and the fine grid yields a maximum deviation below
\(1.7\,\%\). The corresponding particle concentration profiles in
figure~\ref{fig-dns-vs-les-re180-4way-gridstudy-2} reveal a strong
dependence on resolution. The coarsest grid underpredicts the near-wall
concentration by more than an order of magnitude, while the fine grid
closely reproduces the DNS reference with a maximum deviation of
\(35\,\%\). These results highlight the sensitivity of turbophoretic
particle transport to small-scale turbulent motions, consistent with the
findings of \citet{Jantac2024}. Consequently, the \emph{fine} grid is
selected for LES at \(\Rey_\tau = 180\).

At \(\Rey_\tau = 300\), similar trends are observed. The velocity profile
converges towards the DNS with increasing resolution, and the coarsest
grid again overpredicts the bulk velocity, see
figure~\ref{fig-dns-vs-les-re300-4way-gridstudy-1}. From the medium
resolution onwards, good agreement with the in-house DNS is achieved
(within \(1.6\,\%\)). The slightly higher velocities in the logarithmic
region compared with the reference DNS of \citet{Iwamoto2002} are due to
four-way coupling effects, whereby the particle phase transfers momentum
to the fluid. The particle concentration remains more sensitive to grid
resolution than the mean velocity, with the peak concentration differing
by about \(12\,\%\) between the DNS and the fine grid
(figure~\ref{fig-dns-vs-les-re300-4way-gridstudy-2}). Hence, the
\emph{fine} grid is selected for LES at \(\Rey_\tau = 300\).

For higher Reynolds numbers (\(\Rey_\tau = 395\) and \(550\)), the same
qualitative behaviour is observed, though the results become less
sensitive to grid coarsening, see
figure~\ref{fig-dns-vs-les-re395-4way-gridstudy} and
figure~\ref{fig-dns-vs-les-re550-4way-gridstudy}. This reduction in
sensitivity is attributed to the larger outer-scale resolution at higher
Reynolds numbers, where the dominant turbulent structures are already
well captured even on coarser grids. Moreover, LES is inherently more
effective at higher \(\Rey_\tau\) due to the broader range of turbulent
scales. The final grid resolutions used in the Reynolds-number study
presented in Section~\ref{sec-re-study} are summarised in
table~\ref{tbl-les-resolutions}.

\begin{table}
	\begin{center}
		\def~{\hphantom{0}}
		\begin{tabular}{lllll}
			\(\Rey_\tau\) & \(\Delta x^+\) & \(\Delta z^+\) & \(\Delta y^+_w\) & \(\Delta y^+_c\) \\[3pt]
	180 & 9.88 & 6.58 & 0.47 & 6.25 \\
	300 & 9.88 & 6.58 & 0.47 & 6.25 \\
	395 & 9.88 & 6.58 & 0.47 & 6.25 \\
	550 & 19.75 & 6.58 & 0.47 & 6.25 \\
		\end{tabular}
	\caption{Overview of the grid resolution parameters for LES.\label{tbl-les-resolutions}}
	\end{center}
\end{table}

\begin{figure}
	
	\begin{minipage}{0.50\linewidth}
		
		\centering{
			
			\includegraphics[keepaspectratio, width=\textwidth]{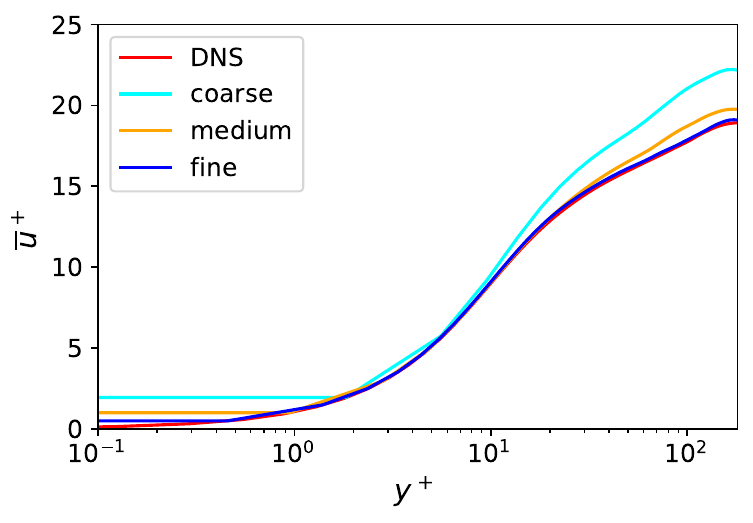}
			
		}
		
		\subcaption{\label{fig-dns-vs-les-re180-4way-gridstudy-1}}
		
	\end{minipage}%
	\begin{minipage}{0.50\linewidth}
		
		\centering{
			
			\includegraphics[keepaspectratio, width=\textwidth]{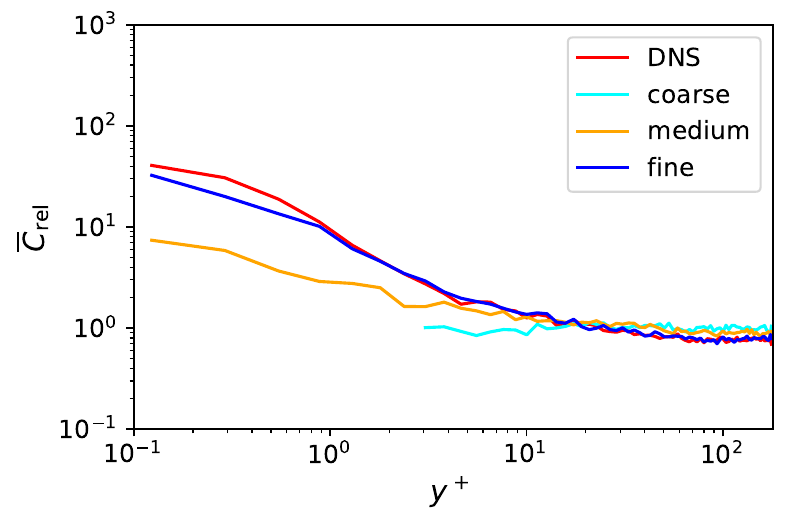}
			
		}
		
		\subcaption{\label{fig-dns-vs-les-re180-4way-gridstudy-2}}
		
	\end{minipage}%
	
	\caption{\label{fig-dns-vs-les-re180-4way-gridstudy}Effect of the LES
		grid resolution on the velocity profile \(\overline{u}^+\) and particle
		concentration \(\overline{C}_\text{rel}\) plotted over wall-normal
		distance \(y^+\). All simulations are conducted at \(\Rey_\tau = 180\)
		with particles of a size with \(d_p = 25 \mu m\). The particles are
		4-way coupled to the fluid.}
	
\end{figure}%

\begin{figure}
	
	\begin{minipage}{0.50\linewidth}
		
		\centering{
			
			\includegraphics[keepaspectratio, width=\textwidth]{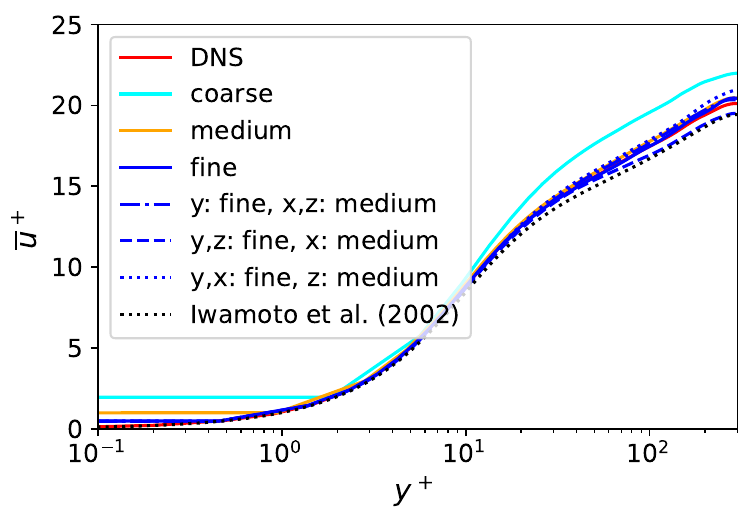}
			
		}
		
		\subcaption{\label{fig-dns-vs-les-re300-4way-gridstudy-1}}
		
	\end{minipage}%
	\begin{minipage}{0.50\linewidth}
		
		\centering{
			
			\includegraphics[keepaspectratio, width=\textwidth]{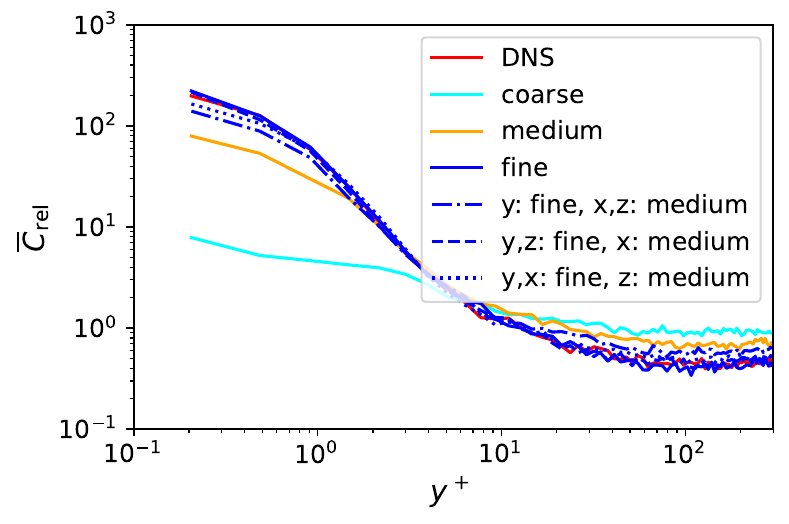}
			
		}
		
		\subcaption{\label{fig-dns-vs-les-re300-4way-gridstudy-2}}
		
	\end{minipage}%
	
	\caption{\label{fig-dns-vs-les-re300-4way-gridstudy}Effect of the LES
		grid resolution on the velocity profile \(\overline{u}^+\) and particle
		concentration \(\overline{C}_\text{rel}\) plotted over wall-normal
		distance \(y^+\). All simulations are conducted at \(\Rey_\tau = 300\)
		with particles of a size with \(d_p = 25 \mu m\). The particles are
		4-way coupled to the fluid.}
	
\end{figure}%

\begin{figure}
	
	\begin{minipage}{0.50\linewidth}
		
		\centering{
			
			\includegraphics[keepaspectratio, width=\textwidth]{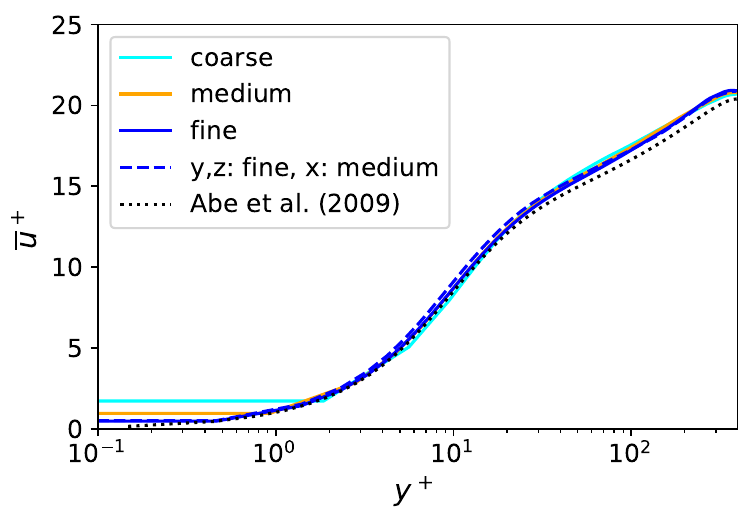}
			
		}
		
		\subcaption{\label{fig-dns-vs-les-re395-4way-gridstudy-1}}
		
	\end{minipage}%
	\begin{minipage}{0.50\linewidth}
		
		\centering{
			
			\includegraphics[keepaspectratio, width=\textwidth]{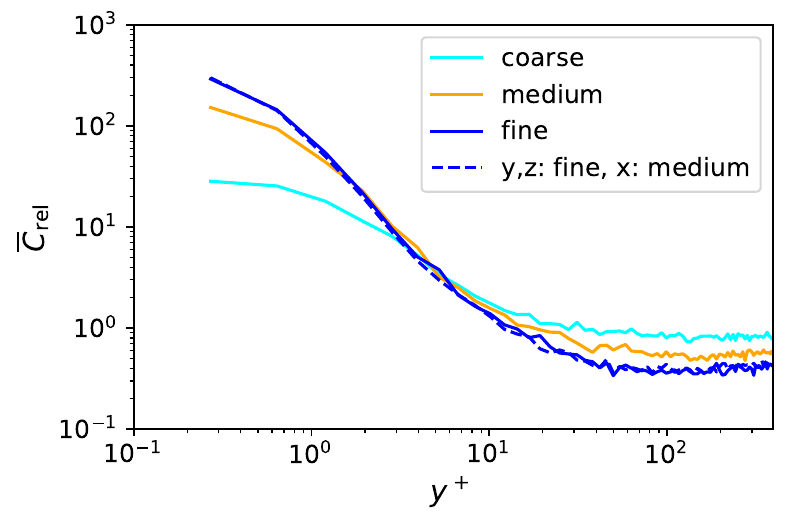}
			
		}
		
		\subcaption{\label{fig-dns-vs-les-re395-4way-gridstudy-2}}
		
	\end{minipage}%
	
	\caption{\label{fig-dns-vs-les-re395-4way-gridstudy}Effect of the LES
		grid resolution on the velocity profile \(\overline{u}^+\) and particle
		concentration \(\overline{C}_\text{rel}\) plotted over wall-normal
		distance \(y^+\). All simulations are conducted at \(\Rey_\tau = 395\)
		with particles of a size with \(d_p = 25 \mu m\). The particles are
		4-way coupled to the fluid.}
	
\end{figure}%

\begin{figure}
	
	\begin{minipage}{0.50\linewidth}
		
		\centering{
			
		\includegraphics[keepaspectratio, width=\textwidth]{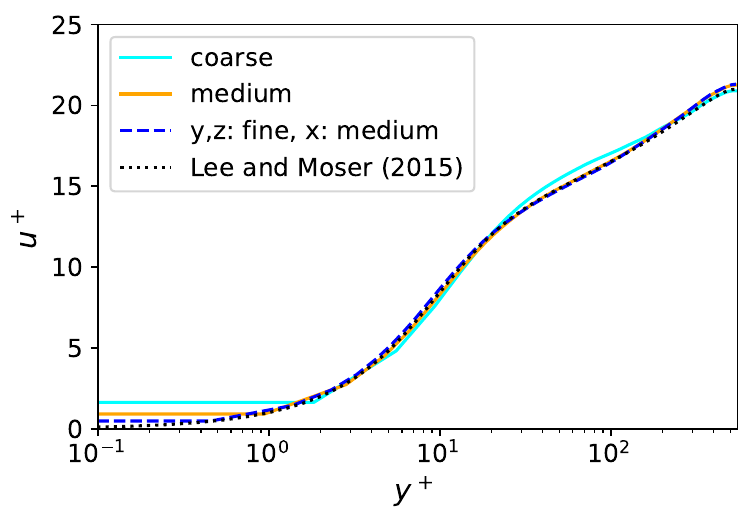}
			
		}
		
		\subcaption{\label{fig-dns-vs-les-re550-4way-gridstudy-1}}
		
	\end{minipage}%
	\begin{minipage}{0.50\linewidth}
		
		\centering{
			
			\includegraphics[keepaspectratio, width=\textwidth]{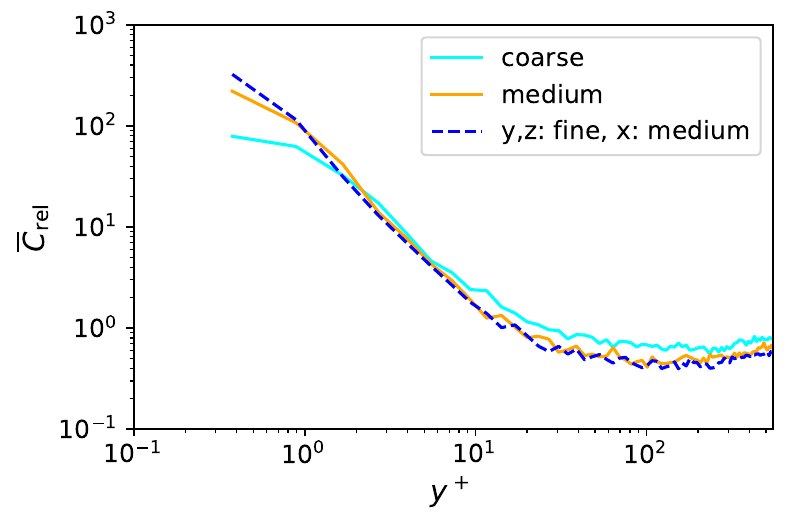}
			
		}
		
		\subcaption{\label{fig-dns-vs-les-re550-4way-gridstudy-2}}
		
	\end{minipage}%
	
	\caption{\label{fig-dns-vs-les-re550-4way-gridstudy}Effect of the LES
		grid resolution on the velocity profile \(\overline{u}^+\) and particle
		concentration \(\overline{C}_\text{rel}\) plotted over wall-normal
		distance \(y^+\). All simulations are conducted at \(\Rey_\tau = 550\)
		with particles of a size with \(d_p = 25 \mu m\). The particles are
		4-way coupled to the fluid.}
	
\end{figure}%

\section{Comparison of condenser and stochastic scaling
	model}\label{sec-apx-condenser-vs-ssm}

The SSM is able to predict a very similar charging behaviour as the
condenser model by setting the parameters of the reference impact
accordingly. Figure~\ref{fig-ssm-vs-condenser} depicts the particle
charge plotted over the number of particle-wall collisions for both
charging models and the three different particle sizes. Up to
approximatly \(10^4\) particle-wall collisions the models do not differ
in the charge built-up. For the investigated time duration, no particles
exceed that number of particle wall-collisions meaning that both
charging models can predict the same charging behaviour. To obtain these
results, the quantities of the reference impact need to be set in the
following way: \(\sigma_0 = \gamma_0 = 0\) and
\(\mu_0 = \Delta q_{0,\text{min}} = \mu_{\text{condenser}}\).
\(\mu_{\text{condenser}}\) is a function of \(u_n\) and \(d_p\) where
\(u_n = 0.01\, \text{m}\,\text{s}^{-1}\) is a matches the average
wall-normal particle impact velocity. For this constellation, particles
of \(d_p = 25\,\unit{\um}\) require
\(\mu_{\text{condenser}} = 2.7702\cdot 10^{-10}\,\text{C}\),
\(d_p = 50\,\unit{\um}\) require
\(\mu_{\text{condenser}} = 2.1417\cdot 10^{-13}\,\text{C}\) and
\(d_p = 100\,\unit{\um}\) require
\(\mu_{\text{condenser}} = 7.7107\cdot 10^{-15}\,\text{C}\).

\begin{figure}
	
		\centering{
		
		\begin{minipage}{0.50\linewidth}
			
			\centering{
			
			\includegraphics[keepaspectratio, width=\textwidth]{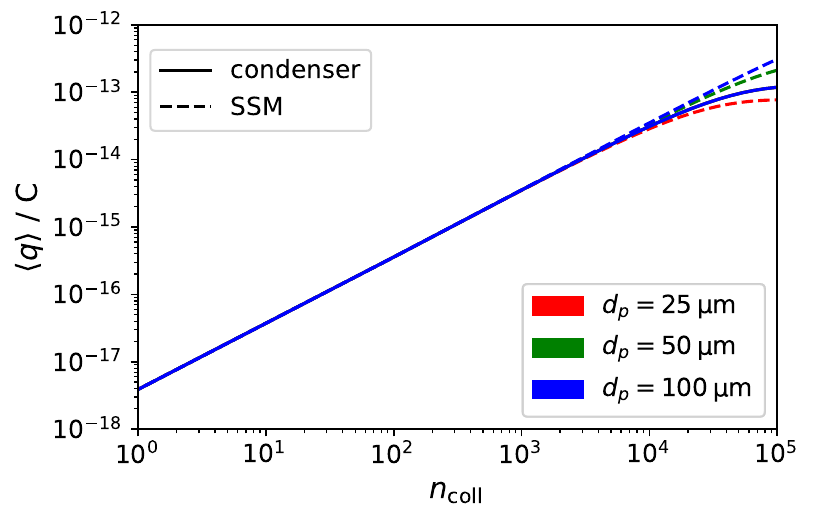}
			
		}
			
		\end{minipage}%
		
	}
	
	\caption{\label{fig-ssm-vs-condenser}Comparison of the condenser model
		with a saturation charge of
		\(q_\text{sat} = 1.23 \cdot 10^{13} \, \text{C}\) and the SSM with
		\(\sigma_0 = \gamma_0 = 0\) to reproduce the condenser model by setting
		\(\mu_0 = \Delta q_{0,\text{min}} = \mu_{\text{condenser}}\). Plotted is
		the particle charge over the number of particle-wall collisions.}
	
\end{figure}%

\end{appen}\clearpage

\nocite{Iwamoto2002, Abe2009, Lee2015}

\end{document}